\documentclass[11pt,a4paper]{article}
\usepackage{jheppub}
\usepackage{amssymb, amsmath, bbm}
\usepackage{graphicx}

\def \nn {\nonumber}
\def \mc {\mathcal}
\def \mbb {\mathbb}
\newcommand\pd{\partial}
\newcommand\pdb{\bar{\partial}}

\newcommand\zb{\bar{z}}
\newcommand\Zb{\bar{Z}}
\newcommand\wb{\bar{w}}
\newcommand\tw{\sigma_{\theta_i}}
\newcommand\la{\langle}
\newcommand\ra{\rangle}
\newcommand\be{\begin{equation}}
\newcommand\ee{\end{equation}}
\newcommand{\bea}{\begin{eqnarray}}
\newcommand{\eea}{\end{eqnarray}}
\newcommand{\ti}{\times}
\newcommand{\half}{\frac{1}{2}}

\title{Scattering and Sequestering of Blow-Up Moduli in Local String Models}
\author{Joseph P. Conlon,}
\author{Lukas T. Witkowski}
\subheader{OUTP/11-51P}
\affiliation{Rudolf Peierls Centre for Theoretical Physics,\\ 1 Keble Road, Oxford OX1 3NP, UK}

\emailAdd{j.conlon1@physics.ox.ac.uk}
\emailAdd{l.witkowski1@physics.ox.ac.uk}

\abstract{We study the scattering and sequestering of blow-up fields - either local to or distant from a visible matter sector -
through a CFT computation of the
dependence of physical Yukawa couplings on the blow-up moduli.
For a visible sector of D3-branes on orbifold singularities we compute the
disk correlator $\langle \tau_s^{(1)} \tau_s^{(2)} \ldots \tau_s^{(n)} \psi \psi \phi \rangle$
between orbifold blow-up moduli and matter Yukawa couplings.
For $n=1$ we determine the full quantum and classical correlator. This result has the correct factorisation
onto lower 3-point functions and also passes numerous other consistency checks.
For $n > 1$ we show that the structure of picture-changing applied to the twist operators
establishes the sequestering of distant blow-up moduli at disk level to all orders in $\alpha'$.
We explain how these results are relevant to suppressing soft terms to scales parametrically below the gravitino mass.
By giving vevs to the blow-up fields we can move into the smooth limit and thereby
derive CFT results for the smooth Swiss-cheese Calabi-Yaus that appear in the Large Volume Scenario.
}

\begin{document}
\maketitle
\section{Introduction and Motivation}

Low energy supersymmetry and its breaking are one of the most promising phenomenological applications for string theory.
If low-energy supersymmetry is realised in nature, then irrespective of the mediation mechanism,
ultraviolet physics will be necessary to understand the structure of the low energy Lagrangian.
For gauge mediation, ultraviolet
physics is necessary to understand the required unnatural lightness of the gravitino mass and the resolution of the problems associated with
the many light moduli that should be present.
For gravity mediation, ultraviolet physics (Planck-suppressed operators) is responsible
for stabilising the weak scale, generating the soft Lagrangian and solving the flavour problem. At the time of writing, the existence of
low-energy supersymmetry is unclear. The current absence of any LHC signals for new physics tells us that the hierarchy problem remains,
but does not yet guide us as to the form of the solution.

The procedure for computing gravity-mediated soft terms is well known \cite{BrignoleIbanezMunoz}.
The first step is the determination of the supergravity Lagrangian, specified by
\bea
\label{sugra}
K & = & \hat{K}(\Phi, \bar{\Phi}) + Z(\Phi, \bar{\Phi})_{ij} C^i \bar{C}^j + \ldots, \\
W & = & W(\Phi) + Y_{\alpha \beta \gamma}(\Phi) C^{\alpha} C^{\beta} C^{\gamma} + \ldots, \\
\label{sugra3}
f_a & = & f_a(\Phi).
\eea
Here $\Phi$ represent moduli fields. These are uncharged, naturally have large vevs and are expected to break supersymmetry.
$C^i$ represent the matter fields of the MSSM (or its extensions). The minimum of the moduli potential is found from the F-term potential
\be
V_F = e^{\hat{K}} \left[ \hat{K}^{i \bar{j}} D_i W D_{\bar{j}} \bar{W} - 3 \vert W \vert^2 \right].
\ee
Using the F-terms of the moduli \emph{in vacuo}, the visible sector soft terms are determined through the standard formulae
\bea
\label{bimsoft}
m_{\alpha}^2 & = & m_{3/2}^2 - F^{\bar{m}} F^n \partial_{\bar{m}} \partial_n \log Z_{\alpha}, \\
A_{\alpha \beta \gamma} & = & F^m \left[ \hat{K}_m + \partial_m Y_{\alpha \beta \gamma} - \partial_m \ln (Z_{\alpha} Z_{\beta} Z_{\gamma}) \right], \\
M_a & = & F^m \frac{ \partial_m f_a}{2 \hbox{Re} f_a}.
\eea
For non-diagonal matter metrics $Z$ the full formulae can be found in e.g. \cite{BrignoleIbanezMunoz}.
The study of supersymmetry breaking in string theory requires extracting the coefficients $\hat{K}$, $Z$, $W$,$Y_{\alpha \beta \gamma}$ and
$f_a$ of (\ref{sugra}) to (\ref{sugra3}) from high-scale
string compactifications.

It is well known to every phenomenologist (and at times repeated \emph{ad nauseam}) that anarchic, unstructured forms
of $Z_{i\bar{j}}(\Phi, \bar{\Phi})$ and $Y_{\alpha \beta \gamma}(\Phi)$ result in soft terms that cannot solve the hierarchy problem: if the soft terms
are at the TeV scale, they are excluded by precision flavour and FCNC constraints. The \emph{locus classicus} for this is
 $K \bar{K}$ mixing, but the constraints arise from many other precision flavour measurements and limit anarchic soft terms to mass
 scales $m \gtrsim 10^3 \hbox{TeV}$.

Low energy actions have been extensively studied in string theory (e.g. see \cite{Dixon1989, KaplunovskyLouis, 0403067, 0404134, 0610327, 0610129, 11093192}).
These studies have revealed remarkable structure in the form of (\ref{sugra}) to (\ref{sugra3}) that belies any naive expectations
of flavour anarchy. Although this structure is natural within string theory, it is not such as would be written down
easily by an effective field theorist. For example, the K\"ahler potential $\hat{K}(\Phi, \bar{\Phi})$ factorises into parts
depending on K\"ahler and complex structure moduli, with corrections suppressed by either the $\alpha'$ or $g_s$ expansion.
The dependence of the matter metric $Z_{i\bar{j}}(\Phi, \bar{\Phi})$ is often universal at leading order and in some cases to
all orders in $\alpha'$ \cite{07052366, 07100873}.

One area where the special structure of (\ref{sugra}) to (\ref{sugra3}) has been particularly well studied is that
of flux compactifications of IIB string theory in the large volume limit. At leading order in the $\alpha'$ expansion, the dilaton and complex structure moduli are fixed supersymmetrically
by fluxes. The residual theory of the K\"ahler moduli corresponds to no-scale supersymmetry breaking,
with the moduli sector appearing as
\bea
\hat{K} & = & - 2 \ln \mc{V}(T + \bar{T}), \nn \\
W & = & W_0.
\eea
Independent of the number of moduli and the form of $\mc{V}(T + \bar{T})$, the susy breaking aligns along the rescaling
direction $g \to \lambda g$.
The Large Volume Scenario (LVS) \cite{0502058, 0505076}, which stabilises the K\"ahler moduli while breaking supersymmetry at hierarchically small
energies, mostly inherits this no-scale structure.
No-scale supersymmetry breaking has many striking features, not least of which are the
vanishing of both tree-level soft terms and loop level anomaly-mediated soft terms \cite{0610129, 9911029, 0004170} .

This leads to a vanishing of soft terms at order both $m_{3/2} \sim \frac{M_P}{\mc{V}}$ and $\frac{\alpha m_{3/2}}{4 \pi}$.
Soft terms have to be generated at some scale, and this
opens the possibility of a spectrum with soft terms parametrically suppressed by powers of $\mc{V} \gg 1$ compared to the gravitino mass
$m_{3/2}$ \cite{09063297, 09122950} (for possibilities of avoiding this conclusion via 1-loop field redefinitions see \cite{10030388, 10110999, 11085740}). However determining exactly which scale soft terms will arise at requires the consideration of higher order corrections both in
$\alpha'$ and $g_s$.

This suppression of soft terms compared to the gravitino mass
is related to the field theory concept of sequestering (for string discussions see \cite{0703105} and in particular \cite{10121858}).
Although no-scale flux compactifications and LVS do not achieve full sequestering, they do satisfy a sort-of
sequestering (discussed in \cite{09063297} and borrowing the name from \cite{10121858}) through the condition
\be
\label{ggg1}
Z(T, \bar{T}) \sim e^{K(T, \bar{T})/3},
\ee
which holds at leading order. The significance of this condition \cite{09063297} is seen by recalling that physical Yukawas $\hat{Y}_{\alpha \beta \gamma}$ are given by
\be
\hat{Y}_{\alpha \beta \gamma} = e^{K/2} \frac{Y_{\alpha \beta \gamma}}{\sqrt{Z_{\alpha} Z_{\beta} Z_{\gamma}}}.
\ee
As the K\"ahler moduli do not appear perturbatively in the holomorphic Yukawa couplings $Y_{\alpha \beta \gamma}$
(\ref{ggg1}) implies that the physical Yukawa couplings $\hat{Y}_{\alpha \beta \gamma}$
do not depend on some of the K\"ahler moduli. For the cases where only the $T$ fields break supersymmetry,
the condition (\ref{ggg1}) leads to exact cancellations in the soft term expressions (\ref{bimsoft}) to give vanishing soft terms.

Sort-of sequestering is known to be present at leading order. However when subleading corrections to the K\"ahler potential are considered
- for example the $\alpha'^3$ term that plays a prominent role in constructing the LVS - it is unclear what happens.
This limited the ability of \cite{09063297} to perform a definite calculation of the scale of volume-suppressed
soft terms. This paper used the subleading correction to the moduli K\"ahler potential, but was ignorant as to the form of subleading
corrections to the matter metrics.

One of the prime aims of this paper is to extend the study of sort-of sequestering beyond leading order to obtain results
at all orders at $\alpha'$, although remaining at leading order in $g_s$. We will do this by working in toroidal orbifold CFT and
computing disk correlators between K\"ahler moduli and matter Yukawa couplings.

\begin{figure}
\begin{center}
\includegraphics[height=5cm]{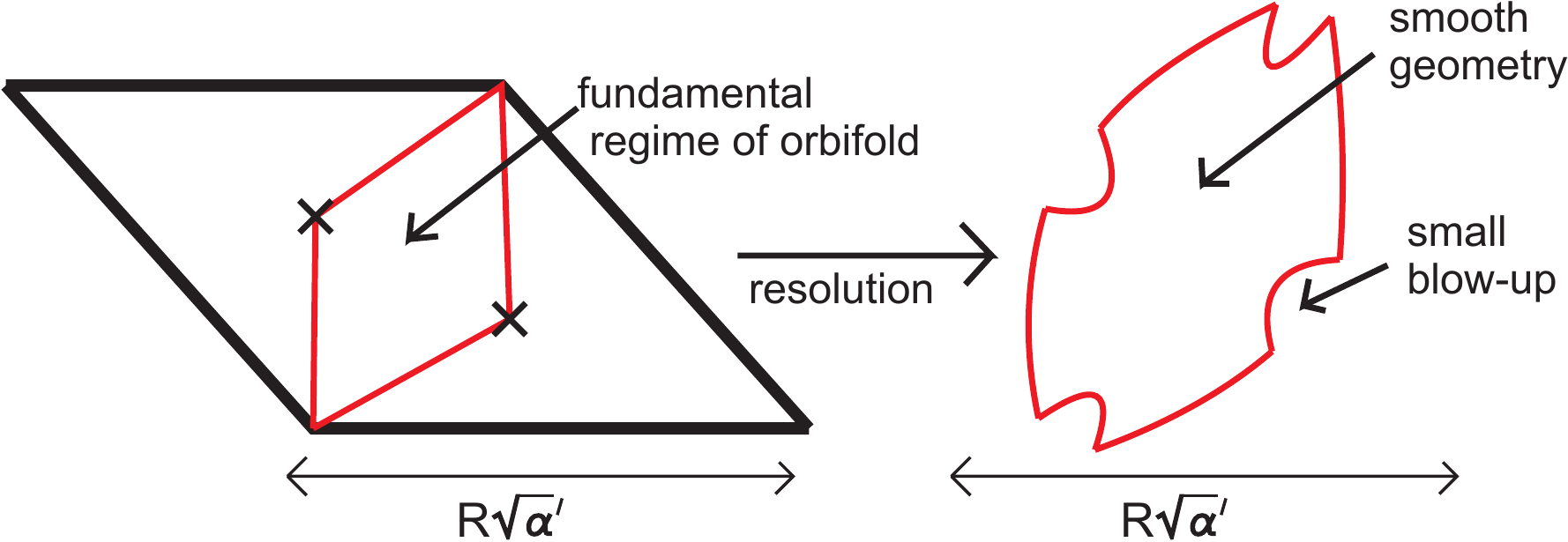}
\caption{Resolving toroidal orbifold singularities gives the type of smooth geometry that appears in the Large Volume Scenario,
with a large bulk attached to several small blow-ups.}
\label{OrbifoldSmooth}\end{center}\end{figure}

A bonus is that
toroidal orbifolds provide a good approximation for the Swiss-cheese Calabi-Yaus
present in the LVS. They consist of a large bulk (parametrised by the three
2-tori) together with many small blow-up cycles, which in the orbifold limit are blown down at the singularities.
By increasing the radii of the tori it is possible to make the bulk arbitrarily large while retaining the exact worldsheet description.
Placing D3-branes at certain singlularities allows the incorporation of semi-realistic matter sectors such as considered in
\cite{aiqu, 0508089, 08105660, 09103616, 10021790, 11066039}.
If the blow-up cycles are resolved, we obtain the typical smooth Calabi-Yau geometry that appears in the LVS, with
a very large bulk and small blow-up cycles.
From a CFT perspective, these smooth Calabi-Yaus are obtained by vevving the marginal directions parametrised by the
blow-up (twist) fields. Although working on an orbifold we can therefore probe the smooth limit - at least within the radius of convergence
of this expansion - by computing correlators involving many blow-up fields. This is illustrated in figure \ref{OrbifoldSmooth}.

As our stated purpose is to use orbifold computations to obtain information about smooth geometries, it behoves us to be precise
in matching orbifold modes onto K\"ahler moduli in the smooth geometry. For a recent related discussion in the context of the heterotic
string, see \cite{11080667}.

\subsection*{Matching Orbifold Modes to K\"ahler Moduli}

In models with multiple moduli the K\"ahler moduli can be classed into several kinds and
for orbifold models we can clearly distinguish three types of modes. These can also be mapped onto
different types of modes in smooth Calabi-Yau models. We provide examples of each kind.
The map is smoothest in `Swiss cheese' geometries such as occur in the Large Volume Scenario, where we can
break the moduli into large bulk moduli and small blow-up moduli.

First, there are the bulk modes. In orbifold language, these correspond to untwisted modes (often also called the
$\mc{N}=4$ sector). For toroidal orbifolds, these are the metric modes that parametrise the
volumes of the compactification tori, and would also be present in the theory prior to the orbifold
projection. The worldsheet coordinate $X^{\mu}(\tau, \sigma)$ obeys
\be
X^{\mu} (\tau, \sigma + 2 \pi) = X^{\mu}(\tau, \sigma),
\ee
for all spatial directions $\mu$. For smooth Calabi-Yaus these modes correspond to the modes that determine the overall volume, or for fibration
Calabi-Yaus (e.g. K3 fibrations) the modes that determine the volume of the fibre and the volume of the base.
These modes can only be normalised across the entire compact space and do not localise in the geometry.

Second, there are the entirely local modes. In orbifold language, these correspond to the fully twisted sector
(also called the $\mc{N}=1$ sector). These modes correspond to local blow-up modes that are located at orbifold
singularities. In the vicinity of the singularity the complexified worldsheet coordinate $X^i(\tau, \sigma)$ obeys
\bea
X^i (\tau, \sigma + 2 \pi) & = & e^{2 \pi i \theta_i} X^i(\tau, \sigma), \nn \\
\bar{X}^{i} (\tau, \sigma + 2 \pi) & = & e^{-2 \pi i \theta_i} \bar{X}^i(\tau, \sigma).
\eea
The $\mc{N}=1$ nature is associated to the requirement that $\theta_1, \theta_2, \theta_3 \neq 0$ (so the twisting is in all compact
directions) and the requirement $\theta_1 + \theta_2 + \theta_3 \in \mbb{Z}$ (to preserve $\mc{N}=1$ supersymmetry).
If the singularity is resolved into a smooth Calabi-Yau these modes map onto localised 2/4-cycles.
We can consider the geometry defined by a non-compact Calabi-Yau attached to the resolved singularity.
Such $\mc{N}=1$ modes map onto localised 2/4-cycles which are normalised in this non-compact geometry. The metric
wavefunction vanishes at infinity, both the
cycle and its dual cycle are defined in the local geometry, and any homology tadpole that is sourced along this cycle must be cancelled locally:
there are no distant `cycles at infinity' in the same homology class.

We enumerate some examples of $\mc{N}=1$ modes.
\begin{enumerate}
\item The local geometry $\mbb{C}^3/\mbb{Z}_3$ (found for example in the Z manifold $T^6/\mbb{Z}_3$)
has two $\mc{N}=1$ twisted modes at the singularity. This can be resolved into the smooth non-compact
Calabi-Yau $\mc{O}_{\mbb{P}^2}(-3)$ for which the explicit metric is known \cite{Lutken}.
The orbifold twisted mode corresponds in the smooth geometry
to the $\mbb{P}^2$ (and its dual cycle) that are present. These metric modes are explicitly normalisable in the resolved geometry.
\item
The non-abelian orbifold $\mbb{C}^3/\Delta_{27}$ gives an orbifold description of the del Pezzo 8 singularity \cite{0508089}.
The string spectrum for the orbifold has two $\mc{N}=1$ twisted states and eight $\mc{N}=2$ twisted modes. In the smooth
resolution of this singularity, the $\mc{N}=1$ modes correspond to the modulus controlling $\hbox{Vol}(dP_8)$
and its local dual 2-cycle, which are
the only modes that unambiguously belong to the homology of the smooth Calabi-Yau.
\item
In the `Swiss cheese' model $\mbb{P}^4_{[1,1,1,6,9]}$, the small cycle is locally described as a $\mbb{C}^3/\mbb{Z}_3$ singularity.
As above, this identifies the small cycle $\tau_s$ that appears in the Large Volume Scenario as the smooth counterpart of an
orbifold $\mc{N}=1$ twisted mode.

\end{enumerate}

The third type of mode are the partially twisted modes (often called the $\mc{N}=2$ twisted sector). These modes
are twisted along two of the directions, and untwisted along the third. The worldsheet obeys
\bea
X^{i}(\tau, \sigma + 2 \pi) & = & e^{2 \pi i \theta_i} X^i(\tau, \sigma), \qquad i = 1,2, \nn \\
X^3(\tau, \sigma + 2 \pi) & = & X^3(\tau, \sigma).
\eea
It is also necessary that $\theta_1 + \theta_2 \in \mbb{Z}$ (this is the condition required to preserve $\mc{N}=2$ supersymmetry).
In the orbifold these modes are tied to the singularity along two of the tori and are free to propagate in the third torus.
Under the orbifold resolution to a smooth space, these correspond to metric modes that are visible locally, but are not normalised
locally.
Consider the geometry defined by the non-compact Calabi-Yau attached to the resolved singularity.
$\mc{N}=2$ modes map onto localised 2-cycles (or \emph{mutatis mutandis} 4-cycles)
for which the dual 4-cycle is non-compact and cannot be defined in the local
geometry. In contrast to $\mc{N}=1$ modes, the metric wavefunction does not vanish at infinity.
An `$\mc{N}=2$ cycle' is defined in the local geometry, and a homology tadpole can be induced locally on this cycle, but the tadpole does not need
to be cancelled locally. In contrast to $\mc{N}=1$ modes, there may be distant `cycles at infinity' which are in the same homology class and on which branes can wrap
cancelling the local tadpole. Alternatively, cycles which appear distinct in the local geometry may be revealed to be homologically identical
in the global geometry.

Let us enumerate some examples of $\mc{N}=2$ twisted modes:
\begin{enumerate}
\item
The toroidal orbifold $T^6/\mbb{Z}_4$ has $\mc{N}=2$ modes located at each of the $\mbb{C}^3/\mbb{Z}_4$ orbifold fixed points.
When examining the geometry of the resolved space, we see that the `$\mc{N}=2$ cycles' in the smooth space are a conjunction of
$\mc{N}=2$ twisted modes that are located at four separate $\mbb{C}^3/\mbb{Z}_4$ singularities. The four local `$\mc{N}=2$ cycles' turn out to be
identical in homology in the global space, and so part of the same cycle, even though they are geographically separated. For further discussion
of $\mc{N}=2$ modes in $T^6/\mbb{Z}_4$, consult \cite{09014350, 09061920, 10084361}.

\item
As mentioned above the non-abelian orbifold $\mbb{C}^3/\Delta_{27}$ describes the del Pezzo 8 singularity at one point in its complex structure.
The orbifold spectrum has eight $\mc{N}=2$ twisted modes. $dP_8$ has nine 2-cycles. One is dual to the overall del Pezzo 4-cycle,
and the eight $\mc{N}=2$ modes correspond to the remaining eight 2-cycles. As is known, (see e.g. \cite{0610007})
in a smooth Calabi-Yau resolution of the $dP_8$ singularity anywhere  these 2-cycles may or may not correspond to
actual modes of the Calabi-Yau; it requires the global geometry. Furthermore, the 2-cycles of the del Pezzo can in fact be homologically identical in the full space.

\item
The conifold (a cone over $T^{1,1}$, obtained from $S^5/\mbb{Z}^2$ as a relevant deformation) has a local $\mbb{P}^1$ 2-cycle.
However in global embeddings there may be other cycles in the same homology class as this $\mbb{P}^1$ (for example see the discussion
of the conifold transition in \cite{9702155}), and tadpoles defined on this $\mbb{P}^1$ do not need to be cancelled locally.
This identified the $\mbb{P}^1$ present in the conifold as having the appropriate properties of an $\mc{N}=2$ mode.
\end{enumerate}

In some ways the $\mc{N}=2$ modes are the trickiest to deal with, as they are neither fully local nor fully global.
For example, they can couple to tadpoles sourced by Standard Model fields even though the appropriate twist operator is at a
singularity far removed from the Standard Model branes.
In this paper we will focus on the couplings of $\mc{N}=1$ modes and analyse their coupling to the visible sector, and leave the
study of $\mc{N}=2$ modes to future work. In the context of the Large Volume Scenario the $\mc{N}=1$ modes
correspond to the `holes in the cheese', and are definitely present and definitely have non-zero F-terms.

In this paper we consider correlators of the form
\be
\langle \tau_s^{(1)} \tau_s^{(2)} \ldots \tau_s^{(n)} \psi \psi \phi \rangle
\ee
where $\tau_s^{(i)}$ corresponds to a twist field associated to a blow-up mode. We shall focus our attentions on
$\mc{N}=1$ modes where the fields $\tau_s^{(i)}$ correspond to twists $(\theta_1, \theta_2, \theta_3)$ with all $\theta_i \neq 0$.
At zero momentum these correlators directly measure the dependence of the physical Yukawa couplings on the blow-up fields and
thus allow a direct test of `sort-of sequestering'. In the case that $n=1$ we shall determine both the quantum and classical correlator
whereas for $n > 1$ we shall limit ourselves to studying the sequestering of distant fields.

The paper is structured as follows. Section 2 gathers the necessary CFT tools and results that are required to compute the correlators we are interested in. The correlators have two parts, classical and quantum. The classical part is relevant when the correlator requires the worldsheet
to stretch between different points in space time, otherwise the classical part admits a trivial solution. The quantum part is independent of the classical behaviour and in the case where the classical solution is trivial gives the full result. In sections 3 and 4 we focus on
cases where all fields are located at the same point and the classical solution is trivial. In section 3, we calculate 3-point functions
involving two open strings and one closed string twisted modulus. This is mostly a warm-up computation although it will also help
us in section 4 as a check of the lower-point reduction. In section 4 we calculate the 4-point function involving
three open strings and one closed string twisted modulus. There are various technical subtleties associated to this
computation (for example the fixing of the $SL(2, \mbb{R})$ invariance). We explain how to resolve them and describe
 various consistency checks that apply to our final result.

 Section 5 consider correlators involving matter fields at one singularity and twisted moduli at another singularity. Here the classical
 solution from worldsheet stretching forces the amplitude to be exponentially suppressed in the bulk radius unless the twist operators
 approach a factorisation limit. By a careful analysis of the OPEs in the factorisation limit and picture changing, we show that this does
 not lead to any zero momentum correlator of the form $\langle \tau_s \tau_s \tau_s \ldots \psi \psi \phi \rangle$, thereby establishing
 sort-of sequestering for these distant modes. In section 6 we conclude and explain the implications of our results.

\section{CFT Building Blocks}
\label{CFTBasics}

In this section we gather the necessary tools for the CFT computations. Useful references are \cite{Polchinski1, Polchinski2}.

\subsection{Untwisted Fields}

\subsubsection*{Basic Correlators}

We here summarize results for correlators of the worldsheet fields $X$, $\psi$ and $\phi$.
The worldsheet fields can be written as
\bea
X^{\mu}(z_1, \bar{z}_1) & = & X^{\mu}_L(z_1) + X^{\mu}_R(\bar{z}_1), \\
\psi^{\mu}(z_1, \bar{z}_1) & = & \psi^{\mu}_L(z_1) + \psi^{\mu}_R(\bar{z}_1), \\
\phi(z_1, \bar{z}_1) & = & \phi_L(z_1) + \phi_R(\bar{z}_1).
\eea
The holomorphic (and similarly the antiholomorphic) elements have the standard correlation functions
\begin{align}
\langle X^{\mu}_L(z_1) X^{\nu}_L(z_2) \rangle = & \ - \eta^{\mu \nu} \ln (z_1 - z_2), \\
\langle \partial X^{\mu}_L(z_1) \partial X^{\nu}_L(z_2) \rangle = & - \frac{\eta^{\mu \nu}}{(z_1 - z_2)^2}, \\
\langle \psi^{\mu}_L(z_1) \psi^{\nu}_L(z_2) \rangle = & \ \frac{\eta^{\mu \nu}}{z_1 - z_2}, \\
\langle \phi_L(z_1) \phi_L(z_2) \rangle = & \ -\ln (z_1 - z_2) \ .
\end{align}
On the disk it is also necessary to satisfy the boundary conditions $\partial_n X = 0$ (Neumann) or $\partial_t X = 0$ (Dirichlet).
The simplest way to obtain the Green's function in the presence of the boundary conditions is to introduce
cross-correlations between the holomorphic and antiholomorphic fields.
On the disk these cross-correlators are:
\begin{align}
\label{ggg}
\langle X^{\mu}_L(z_1) X^{\nu}_R(\zb_2) \rangle = & \ - D^{\mu \nu} \ln (z_1 - \zb_2), \\
\langle \psi^{\mu}_L(z_1) \psi^{\nu}_R(\zb_2) \rangle = & \ \frac{D^{\mu \nu}}{z_1 - \zb_2}, \\
\langle \phi_L(z_1) \phi_R(\zb_2) \rangle = & \ -\ln (z_1 - \zb_2) \ .
\end{align}
The matrix $D^{\mu \nu}$ encapsulates the conditions on the disk boundary:
\begin{equation}
  D^{\mu \nu} = \left\{
  \begin{array}{l l}
    \hphantom{-} \eta^{\mu \nu} & \quad \textrm{Neumann}\\
    - \eta^{\mu \nu} & \quad \textrm{Dirichlet}
  \end{array} \right. \ .
\end{equation}
The full disk correlator is then given by
\bea
\label{abcd}
\langle X^{\mu}(z_1, \bar{z}_1) X^{\nu} (z_2, \bar{z}_2)  \rangle & = & - \eta^{\mu \nu} \left( \ln(z_1 - z_2) + \ln (\bar{z}_1 - \bar{z}_2)
\pm \ln (z_1 - \bar{z}_2) \pm \ln (\bar{z}_1 - z_2) \right), \nn \\
\langle \psi^{\mu}(z_1, \bar{z}_1) \psi^{\nu} (z_2, \bar{z}_2)  \rangle & = & \eta^{\mu \nu} \left( \frac{1}{z_1 - z_2} + \frac{1}{\bar{z}_1 - \bar{z}_2} \pm \frac{1}{z_1 - \bar{z}_2} \pm \frac{1}{\bar{z}_1 - z_2} \right),
\eea
where plus signs apply for Neumann boundary conditions and minus signs for Dirichlet boundary conditions.

Neumann boundary conditions apply for dimensions parallel to D-branes and Dirichlet boundary conditions
apply for dimensions transverse to the D-branes. The models we consider involve (fractional) D3 branes at orbifold singularities, which necessitate
Neumann boundary conditions for the external directions $\mu, \nu=0,1,2,3$ and Dirichlet boundary conditions for the internal directions $\mu, \nu=4,5,6,7,8,9$.

For the fermionic field $\psi$, the correlator (\ref{abcd}) only applies directly to vertex operators located in the bulk of the space.
For the boundary vertex operators that describe open string fields, it is appropriate to restrict purely to the holomorphic
component (so $\langle \psi^{\mu}(z_1) \psi^{\nu}(z_2) \rangle = \frac{\eta^{\mu \nu}}{z_1 - z_2}$ when $z_1 = \bar{z}_1$ and
$z_2 = \bar{z}_2$).

It is useful to work with a different basis of fields in the
internal dimensions.
In particular, we rewrite the above expressions in terms of complexified internal bosonic coordinates and bosonised internal spinors.
One can complexify the bosonic fields as follows
\begin{equation}
Z_i= \frac{1}{\sqrt{2}}\left(X^{2i+2} + i X^{2i+3} \right), \quad \quad \Zb_i= \frac{1}{\sqrt{2}}\left(X^{2i+2} - i X^{2i+3} \right), \quad \quad i=1,2,3 \ .
\end{equation}
These satisfy
\be
\langle \partial Z_i (z) \partial Z_i(w) \rangle = 0, \qquad \langle \partial Z_i(z) \partial \bar{Z}_i(w) \rangle = - \frac{1}{(z-w)^2}.
\ee
We also form complex combinations of worldsheet spinors:
\begin{equation}
\Psi_i= \frac{1}{\sqrt{2}}\left(\psi^{2i+2} + i \psi^{2i+3} \right), \quad \quad \bar{\Psi}_i= \frac{1}{\sqrt{2}}\left(\psi^{2i+2} - i \psi^{2i+3} \right), \quad \quad i=1,2,3,
\end{equation}
with similar expressions for the antiholomorphic counterparts $\tilde{\Psi}$ and $\tilde{\bar{\Psi}}$.
The complexified spinor can now be bosonised in terms of free scalar fields $H(z)$ as follows:
\begin{equation}
\Psi_i(z) \cong e^{iH_i(z)}, \qquad \bar{\Psi}_i(z) \cong e^{-iH_i(z)} \ .
\end{equation}
We can compute correlation functions of these fields according to
\begin{equation}
\label{spincorr}
\langle \prod_i e^{i q_i \cdot H(z_i)} \rangle = \prod_{i<j} z_{ij}^{q_i \cdot q_j} \ ,	\qquad \sum_i q_i=0 \ .
\end{equation}
The result (\ref{spincorr}) is also valid for amplitudes involving the spin fields $S(z)$
\begin{equation}
S(z)=e^{i \sum_{i=1}^5 q_i H_i(z)} \ , \qquad q=\left(\pm \frac{1}{2}, \pm \frac{1}{2},\pm \frac{1}{2},\pm \frac{1}{2},\pm \frac{1}{2} \right).
\end{equation}

\subsubsection*{Vertex operators}

The string computation proceeds by computing the correlation function of vertex operators inserted on the worldsheet.
For our model with D3-branes at orbifold singularities, the matter states
come from the open string sector while (twisted) moduli arise in the closed string sector.
The open string vertex operators are untwisted and are inserted on the boundary of the disk.
For tree level diagrams, the effect of the orbifold on open string states
only enters explicitly in the requirement that the physical states are invariant under the orbifold projection.

We collect the relevant vertex operators. A space-time gauge boson in the canonical $(-1)$ picture can be written as
\begin{equation}
V_{A^{a}}^{(-1)}(z) = \ \lambda^{a} {\xi}_\mu e^{-\phi} \psi^\mu e^{i k \cdot X}(z),
\end{equation}
where $\lambda^{a}$ is a Chan-Paton factor and ${\xi}_\mu$ encodes the polarization.

Using the complexified worldsheet spinor $\Psi$ a space-time scalar can be written as
\begin{equation}
V_{C_{i}}^{(-1)}(z) = \ \lambda e^{-\phi} \Psi_i e^{i k \cdot X} (z)\ .
\end{equation}
For space-time fermions, the canonical $(-1/2)$ ghost picture vertex operator is given by
\begin{equation}
V_{\psi}^{(-\frac{1}{2})}(z) = \lambda \ e^{-\frac{1}{2} \phi} e^{i q \cdot H} e^{i k \cdot X}(z),
\end{equation}
where all $q_i=\pm \frac{1}{2}$. The GSO projection requires the number of negative entries to be even.

On the disk, the ghost charges of all vertex operator insertions have to sum to $-2$.
This requires picture-changing some of the vertex operators, by
evaluating the limit
\be
V^{(c+1)}(w) = \lim_{z \rightarrow w} e^{\phi(z)} T_{F}(z) V^{(c)}(w),
 \ee
 and discarding terms of $\mathcal{O}{(z-w)}^{-1}$.\footnote{More rigorously, one inserts the picture-changing operator $e^{\phi(z)} T_{F}(z) V^{(c)}(w)$ into the correlation function together with the vertex operators and performs the limit $z \rightarrow w$ only after the calculation of the amplitude. In most cases it is unambiguous to perform the picture-changing first and we will do so wherever possible as it leads to simpler calculations.} The picture-changing operator involves the worldsheet supercurrent $T_F$ which takes the following form on the boundary of the disk:
\begin{equation}
T_F(z) = \pd X_{\mu}\psi^{\mu}(z) + \sum_{i=1}^{3} \left[\pd \Zb_i \Psi_i (z) + \pd Z_i \bar{\Psi}_i(z) \right] \ .
\end{equation}
In principle we could also choose to picture-change the closed string vertex operators. However this involves dealing with excited
twist operators and we do not pursue this option.

To evaluate the picture-changed operators we use the following operator product expansions (OPEs):
\begin{align}
e^{iaH(w)}e^{ibH(z)}= & \ {(w-z)}^{ab} \ e^{i(a+b)H(z)} + \ldots \\
e^{a \phi (w)}e^{b \phi (z)}= & \ {(w-z)}^{-ab} \ e^{(a+b) \phi (z)} + \ldots \\
\pd Z(w)e^{ikX(z)}= & \ - \frac{i}{2} k^{+} {(w-z)}^{-1} e^{ikX(z)} + \pd Z(z)e^{ikX(z)} + \ldots \\
\pd \Zb(w)e^{ikX(z)}= & \ - \frac{i}{2} k^{-} {(w-z)}^{-1} e^{ikX(z)} + \pd Z(z)e^{ikX(z)} + \ldots
\end{align}
The ellipses denote less divergent terms. We have introduced complex momenta $k^{\pm}$, where $k^{\pm} =k^{2n + 1} \pm i k^{2n+2}$ for any complex plane except the $01$ plane, where $k^{\pm}= \pm k^0 +k^1$, and which satisfy
\begin{equation}
k \cdot X(z) \equiv \frac{1}{2} \left(k^{+} \cdot \Zb(z) + k^{-} Z(z) \right) \ .
\end{equation}
Note that for the picture changing of a space time scalar, the $0$-ghost picture operator has the form
\be
V^{(0)}_{C_i} = \lambda \left( \partial_{t(n)} Z_i + i(k \cdot \psi) \psi_i \right) e^{ik \cdot X}(z),
\ee
where the tangential (normal) derivative applies for Neumann (Dirichlet) boundary conditions.

\subsection{Twisted Fields}

We also need to include correlators and vertex operators involving twisted bosonic fields. We thus review some of the basic physics
of twist operators. The technology we will be using was pioneered in \cite{Dixon:1986qv} for twist fields on worldsheets without a boundary. To extend these methods to the disk worldsheet we also follow a suggestion given in the appendix of \cite{Douglas:1996sw}. An alternative approach using cut Riemann surfaces was developed by \cite{Hamidi:1986vh}. In the context of heterotic string theory these methods were employed in \cite{Burwick:1990tu, 9207049, 07114894}. Twisted open strings appear in D-brane models in type I/II string theory and have been studied in \cite{0003185, 0011281, 0303083, 0303124, 0310257}. In the following we generalise these techniques for twisted closed strings on the disk worldsheet. The correlator of two twists in the bulk of the disk has been calculated in \cite{07063199}.

\subsubsection*{Bosonic Twist Fields and Correlators}

Twist operators are used to describe the blow-up fields that are present and localised at orbifold singularities.
A target space orbifold on complexified coordinates $Z_i$ requires us to identify points related by a complex rotation:
$Z_i \sim e^{2 \pi i \theta_i} Z_i$. The closed string Hilbert space is extended by the presence of twisted sectors
defined by $Z_i(\sigma=2 \pi)= e^{2 \pi i \theta_i} Z_i(\sigma = 0)$.
This can be rewritten in terms of the worldsheet coordinates $z, \zb$ as the monodromy condition
$$
Z_i(e^{2 \pi i}z, e^{-2 \pi i} \zb)= e^{2 \pi i \theta_i} Z_i(z, \zb),
$$
where $z, \zb =0$ is a fixed point of the orbifold.
In the worldsheet CFT the monodromy condition is enforced through the
introduction of a bosonic twist field $\tw(z, \zb)$. Generally, there exists
one twist field for each fixed point of the orbifold and hence twist fields
should carry a label for each fixed point.
This affects the classical parts of amplitudes and we defer discussion of this to sections 2.4 and 5.
The monodromy of the coordinate field $Z$ implies the following: in the neighbourhood of a twist field at the origin the field $Z$ undergoes a phase rotation. With this knowledge one can derive the local OPEs for the twist fields \cite{Dixon:1986qv}:
\begin{align}
\label{OPES}
\partial_z Z_i (z, \bar{z}) \ \tw(w, \bar{w}) &\sim {(z-w)}^{-(1-\theta_i)} \ \tau_{\theta_i}(w, \bar{w}) + \cdots \\
\nn \partial_z \bar{Z}_i (z, \bar{z}) \ \tw(w, \bar{w}) &\sim {(z-w)}^{-\theta_i} \ \tau^{\prime}_{\theta_i}(w, \bar{w}) + \cdots \\
\nn \partial_{\zb} Z_i (z, \bar{z}) \ \tw(w, \bar{w}) &\sim {(\zb-\bar{w})}^{-\theta_i} \ \tilde{\tau}^{\prime}_{\theta_i}(w, \bar{w}) + \cdots \\
\nn \partial_{\zb} \bar{Z}_i (z, \bar{z}) \ \tw(w, \bar{w}) &\sim {(\zb-\bar{w})}^{-(1-\theta_i)} \ \tilde{\tau}_{\theta_i}(w, \bar{w}) + \cdots
\end{align}
where we defined four different excited twist fields denoted by $\tau$. Here we require $\theta \in (0,1)$. The analogous OPEs for a twist by a negative angle can be derived from the expressions above. As twisting by $-\theta$ is analogous to a twist by $1-\theta$ we can write down the OPEs for the anti-twist operator $\sigma_{-\theta_i}(w, \bar{w})$ by letting $\theta_i \rightarrow 1-\theta_i$ in (\ref{OPES}):
\begin{align}
\label{OPES2}
\partial_z Z_i (z, \bar{z}) \sigma_{-\theta_i}(w, \bar{w}) &\sim {(z-w)}^{-\theta_i} \ \tau_{-\theta_i}(w, \bar{w}) + \cdots \\
\nn \partial_z \bar{Z}_i (z, \bar{z}) \sigma_{-\theta_i}(w, \bar{w}) &\sim {(z-w)}^{-(1-\theta_i)} \ \tau^{\prime}_{-\theta_i}(w, \bar{w}) + \cdots \\
\nn \partial_{\zb} Z_i (z, \bar{z}) \sigma_{-\theta_i}(w, \bar{w}) &\sim {(\zb-\bar{w})}^{-(1-\theta_i)} \ \tilde{\tau}^{\prime}_{-\theta_i}(w, \bar{w}) + \cdots \\
\nn \partial_{\zb} \bar{Z}_i (z, \bar{z}) \sigma_{-\theta_i}(w, \bar{w}) &\sim {(\zb-\bar{w})}^{-\theta_i} \ \tilde{\tau}_{-\theta_i}(w, \bar{w}) + \cdots \ .
\end{align}
The monodromy conditions lead to cuts $\sim z^{\theta_i}$ in the map $z \rightarrow \partial Z_i(z)$ from the worldsheet to the target space. The exact form of this singular behaviour is encoded in the OPEs when a coordinate field approaches a twist field. These contain enough information to derive correlation functions including the twist fields $\tw(z, \zb)$ on the sphere. To be able to do the same calculations on the disk $D_2$ it is
 necessary to know the cut structure on $D_2$ and we therefore reexamine the above results for the case of the disk.

We will employ the doubling trick to determine the behaviour of twist fields in the bulk of the disk: it will allow us to infer the cut-structure on the disk from that on the sphere. The disk can be conformally mapped to the upper-half plane with the real line as the boundary. The bosonic worldsheet fields $\pd X(z)$ and $\pdb X(\zb)$ thus only have support for $\textrm{Im}(z) \geq 0$ and $\textrm{Im}(\zb) \geq 0$. When going to the target space the boundary has to be mapped to the worldvolume of the D3-branes and hence we need to ensure that the coordinate field $X$ obeys Neumann boundary conditions when $X$ is a space-time coordinate parallel to the D3-branes and Dirichlet boundary conditions otherwise. On the z-plane the boundary conditions acquire the following form:
\begin{equation}
  \partial X = \left\{
  \begin{array}{l l}
    \hphantom{-} \bar{\partial} X & \quad \textrm{Neumann}\\
    - \bar{\partial} X & \quad \textrm{Dirichlet}
  \end{array} \right. \quad \textrm{Im}(z) = \textrm{Im}(\zb) =0 .
\end{equation}
The complexified coordinate fields $Z$ and $\Zb$ both satisfy similar boundary conditions on the real line as they are linear combinations of $X$-fields:
\begin{equation}
\label{BCdisk}
  \partial Z = \left\{
  \begin{array}{l}
    \hphantom{-} \bar{\partial} Z \\
    - \bar{\partial} Z
  \end{array} \right. \quad , \quad
  \partial \Zb = \left\{
  \begin{array}{l l}
    \hphantom{-} \bar{\partial} \Zb & \quad \textrm{Neumann}\\
    - \bar{\partial} \Zb & \quad \textrm{Dirichlet}
  \end{array} \right.
  \quad \textrm{Im}(z) = \textrm{Im}(\zb) =0 .
\end{equation}

Using the above boundary condition the doubling trick then advises us to define new fields $\pd \mathcal{Z}(z)$ and $\pd \bar{\mathcal{Z}}(z)$ that are defined on the whole complex plane by stitching together holomorphic and antiholomorphic fields from the half-plane.
\begin{equation}
\label{doubling}
  \partial \mathcal{Z}(z) = \left\{
  \begin{array}{l l}
    \hphantom{\pm} \pd Z(z) & \quad \textrm{Im}(z) \geq 0\\
    \pm \pdb Z(\zb) & \quad \textrm{Im}(z) < 0
  \end{array} \right. \quad , \quad
\partial \bar{\mathcal{Z}}(z) = \left\{
  \begin{array}{l l}
    \hphantom{\pm} \pd \Zb(z) & \quad \textrm{Im}(z) \geq 0\\
    \pm \pdb \Zb(\zb) & \quad \textrm{Im}(z) < 0
  \end{array} \right.
\end{equation}
The sign encodes the Neumann or Dirichlet boundary conditions. It will be useful for us to define the antiholomorphic fields $\pdb \mathcal{Z}(\zb)$ and $\pdb \bar{\mathcal{Z}}(\zb)$ in a similar fashion, although they will contain the same physical information as the holomorphic ones defined in (\ref{doubling}).

The new fields $\pd \mathcal{Z}(z)$, $\pd \bar{\mathcal{Z}}(z)$, $\pdb \mathcal{Z}(\zb)$ and $\pdb \bar{\mathcal{Z}}(\zb)$ have support on the whole complex plane and one can show that they exhibit the standard correlators for complex bosonic fields on the sphere. By examining the correlation functions of the fields on the sphere one can then infer the results for the fields on the disk. This way one can establish the cross-correlations between holomorphic and anti-holomorphic fields on the disk in (\ref{ggg}). Now we will examine the cut-structure on the disk by applying our results from the sphere. The local behaviour of the field $\pd \mathcal{Z}(z)$ about a twist field is given by (\ref{OPES}): $\pd \mathcal{Z}(z) \ \tw(w, \bar{w}) \sim {(z-w)}^{-(1-\theta_i)} $. Hence we learn:
\begin{equation}
  \pd \mathcal{Z}(z) \ \tw(w, \bar{w}) = \left\{
  \begin{array}{l l l}
    \hphantom{\pm} \pd Z(z) \ \tw(w, \bar{w}) & \sim {(z-w)}^{-(1-\theta_i)}   & \quad \textrm{Im}(z) \geq 0\\
    \pm \pdb Z(\zb) \ \tw(w, \bar{w}) & \sim  {(\zb-w)}^{-(1-\theta_i)}  & \quad \textrm{Im}(z) < 0
  \end{array} \right.
\end{equation}
We discover that the field $\pdb Z(\zb)$ on the disk acquired a monodromy condition about the worldsheet point $w$ which it does not possess on the sphere. As the OPEs on the sphere are local expressions they are still true on the disk. Hence, the field $\pdb Z(\zb)$ will still exhibit the same local behaviour about $\wb$ as on the sphere. By repeating the above analysis for the fields $\pd \bar{\mathcal{Z}}(z)$, $\pdb \mathcal{Z}(\zb)$ and $\pdb \bar{\mathcal{Z}}(\zb)$ one can thus determine the full cut-structure on the disk:
\bea
\label{OPED}
\pd Z_i (z, \bar{z}) \tw(w, \bar{w}) & \sim & \left\{ \begin{array}{lr} (z-w)^{-(1-\theta_i)} & z \rightarrow w \\
(z-\wb)^{-\theta_i} & z \rightarrow \wb \end{array} \right.   \\
\label{OPED1}
\pd \bar{Z}_i(z, \bar{z}) \tw(w, \bar{w}) & \sim & \left\{
\begin{array}{lr}
(z-w)^{-\theta_i} & z \rightarrow w \\
(z-\wb)^{-(1-\theta_i)} & z \rightarrow \wb
\end{array} \right.  \\
\label{OPED2}
\pdb Z_i(z, \bar{z}) \tw(w, \bar{w}) & \sim & \left\{ \begin{array}{lr}
(\zb-\wb)^{-\theta_i} & \zb \rightarrow \wb \\
(\zb-w)^{-(1-\theta_i)} & \zb \rightarrow w
\end{array} \right.  \\
\label{OPED3}
\pdb \bar{Z}_i(z, \bar{z}) \tw(w, \bar{w}) & \sim & \left\{
\begin{array}{lr}
(\zb-\bar{w})^{-(1-\theta_i)} & \zb \rightarrow \wb \\
(\zb-w)^{-\theta_i} & \zb \rightarrow w \ .
\end{array} \right.
\eea

Another way of summarizing the above results is to reinterpret the twist field $\sigma_\theta^{D_2}(w, \wb)$ on the disk. The above asymptotic behaviour implies that the twist field on the disk has the properties of a twist and an anti-twist on the sphere:
\begin{align}
\label{factor}
\sigma_\theta^{D_2}(w, \wb) &\sim \sigma_\theta^{S_2}(w) \ \sigma_{-\theta}^{S_2}(w^{\prime}) \hbox{ where } \ w^{\prime} = \wb, \\
\nn \hphantom{\sigma_\theta^{D_2}(w, \wb)} &\sim \sigma_{-\theta}^{S_2}(\wb^{\prime}) \ \sigma_{\theta}^{S_2}(\wb) \hbox{ where } \ \wb^{\prime} = w,
\end{align}
where $\wb^{\prime}$ and $w^{\prime}$ are treated as (anti-) holomorphic variables on the sphere. It is a matter of taste whether one prefers to work with twist fields on the disk and use the extended singular behaviour (\ref{OPED}-\ref{OPED3}) or one chooses to decompose the twist field as shown above and use the results on the sphere (\ref{OPES}).

Correlation functions involving bosonic twist fields require more thought than correlators with fermionic twist fields and
the general formalism for multi-twist correlators is described in \cite{Dixon:1986qv}.
However in practice we will only encounter either 1-pt amplitudes of pure twist fields $\langle \tw(w, \wb) \rangle$ or the twisted Green's function $\langle \pd Z_i(z_1) \pd \Zb_i(z_2) \tw(w, \wb) \rangle$.\footnote{In the literature the twisted Green's function is correctly defined as $g(z_1,z_2) \equiv - \frac{1}{2} \langle \pd Z(z_1) \pd \Zb(z_2) \sigma_\theta(w, \wb) \rangle / \langle \sigma_\theta(w, \wb) \rangle$. By abuse of notation we will continue to call $\langle \pd Z_i(z_1) \pd \Zb_i(z_2) \tw(w, \wb) \rangle$ a twisted Green's function for the benefit of a definite name.} In both cases the methods developed in \cite{Dixon:1986qv} allow us to derive the necessary expressions.

We begin with correlation functions of twist fields with no other insertions. We first give the correlator of a single
twist field on the disk.
This can be computed by
using a straightforward generalisation to the disk of the stress-tensor method of \cite{Dixon:1986qv} together with the relations (\ref{OPED}-\ref{OPED3}).
As the argument is very similar to that of \cite{Dixon:1986qv} and the result is intuitive we omit the derivation and simply state the results.

Vertex operators for twisted states contain one bosonic twist field for each subtorus in the compact space and hence it is the correlator $\langle \prod_{i=1}^3 \tw(w, \wb)\rangle$ that we wish to evaluate. As the twists on different tori are independent of one another, they are uncorrelated and the amplitude factorizes:
\begin{equation}
\label{self}
\langle \displaystyle\prod_{i=1}^3 \tw (w, \wb) \rangle_{D_2} = (w-\wb)^{-\sum_{i=1}^3\theta_i(1-\theta_i)} .
\end{equation}
Note that on the sphere the correlation function of a single twist operator vanishes. On the disk however, the boundary conditions allow
a contraction between the holomorphic and antiholomorphic components of the twist operator. Eq. (\ref{self}) is in effect the sphere correlator
$\langle \sigma_{\theta}^{S_2} (w) \sigma_{-\theta}^{S_2} (w^{\prime}) \rangle$ where $w^{\prime}=\bar{w}$.

We next present the result for the twisted Green's function $\langle \pd Z_i(z_1) \pd \Zb_i(z_2) \tw(w, \wb) \rangle$ where the index $i$ labels the $i$-th subtorus $\mathbb{T}_i^2$ of the compact space. The coordinate fields parameterize the same subtorus which supports the twist.
 This twisted Green's function is completely determined by the OPEs and is given by:
\begin{align}
\label{Gtw}
\nn \langle \pd Z_i(z_1) \pd \Zb_i(z_2) \tw(w, \wb) \rangle_{D_2} = \ & -(z_1 - z_2)^{-2} \ (w-\wb)^{-\theta_i(1-\theta_i)} \\
\nn & (z_1 - w)^{-(1-\theta_i)} \ (z_1 - \wb)^{-\theta_i} \ (z_2 - w)^{-\theta_i} \ (z_2 - \wb)^{-(1-\theta_i)} \\
& \left[(z_1-w)(z_2-\wb) - \theta_i (z_1-z_2)(w-\wb) \right] .
\end{align}
The first line of the expression has the characteristic double pole of a Green's function $\langle \partial Z \partial \bar{Z} \rangle$
as $z_1 \rightarrow z_2$ and the self-contraction of the twist field. The second line is determined by the singular behaviour (\ref{OPED}) to
(\ref{OPED3}) when a coordinate field approaches the position of a twist. The factor in angular brackets ensures
there are no simple poles as $z_1 \rightarrow z_2$.

Combining (\ref{self}) and (\ref{Gtw}) we arrive at the following expression:
\begin{align}
\label{ZZtw}
\nn \langle \pd Z_i(z_1) \pd \Zb_i(z_2) \displaystyle\prod_{j=1}^3 \sigma_{\theta_j}(w, \wb) \rangle_{D_2} = \ &-(z_1 - z_2)^{-2} \ (w-\wb)^{-\sum_{j=1}^3\theta_j(1-\theta_j)} \\
\nn & (z_1 - w)^{-(1-\theta_i)} \ (z_1 - \wb)^{-\theta_i} \ (z_2 - w)^{-\theta_i} \ (z_2 - \wb)^{-(1-\theta_i)} \\
& \left[(z_1-w)(z_2-\wb) - \theta_i (z_1-z_2)(w-\wb) \right] \ .
\end{align}
There is actually a subtlety in (\ref{ZZtw}) that however does not affect the actual result. The vertex operators for fields with
 Dirichlet boundary conditions are not $\partial Z_i$ but instead $\partial_n Z_i$, the derivative normal to the boundary.
 Writing $z = x + i y$, $\partial_n \equiv \partial_y = i(\partial - \bar{\partial})$. So the correlator we actually require is
 not (\ref{ZZtw}) but
 \be
 \label{jofk}
 \left\langle (\partial - \bar{\partial}) Z_i (z_1) (\partial - \bar{\partial}) \bar{Z}_i(z_2) \sigma_{\theta}(w) \sigma_{-\theta}(\bar{w}) \right\rangle.
 \ee
 In fact, each of the four subcorrelators that enter (\ref{jofk}) give identical results, and we recover (\ref{ZZtw}). To see this, note
 that when $z = \bar{z}$, the local OPEs of $\partial Z (z) \sigma_{\theta}(w,\bar{w})$ and $\bar{\partial} Z (\bar{z})
 \sigma_{\theta}(w,\bar{w})$ are identical (using (\ref{OPED}) and (\ref{OPED2})). This ensures the singular
 behaviour is identical for all terms in (\ref{jofk}).
 Furthermore, the relative minus sign between
 $\langle \partial Z \partial \bar{Z} \rangle$ and $\langle \partial Z \bar{\partial} \bar{Z} \rangle$
 correlators in (\ref{jofk}) is cancelled by the relative minus sign in the
 Dirichlet correlator (\ref{abcd}).

\subsubsection*{Fermionic Twist Fields and Correlators}

Fermionic twist fields are easy to describe using the techniques of bosonisation.
Analogous to the bosonic case one introduces fermionic twist fields $s_{\theta_i}(w)$ and $\tilde{s}_{\theta_i}(\wb)$ such that the worldsheet spinors pick up a phase when approaching the insertion point of a twist. To get a consistent theory one defines the phase such that the worldsheet supercurrents $T_F(z) = \partial X \cdot \psi$ and $\tilde{T}_F(\zb) = \bar{\partial} X \cdot \tilde{\psi}$ are single-valued around the insertion point of a twist $s_{\theta_i}(w) \tilde{s}_{\theta_i}(\wb) \tw(w, \wb)$.  From the OPEs, this is true if the fermionic twist fields are bosonised as
\begin{align}
s_{\pm \theta_i}(w) &= e^{\pm i \theta_i H_i(w)}, \\
\nn \tilde{s}_{\pm \theta_i}(\wb) &= e^{\mp i \theta_i \tilde{H}_i(\wb)},
\end{align}
where $H$ and $\tilde{H}$ are free scalar fields. Furthermore, this is correct for both the sphere and the disk in virtue of the bosonisation in terms of $H$ and $\tilde{H}$.

Analogous to the bosonic self-contraction (\ref{self}), there is a non-zero self contraction on the disk, when
we in effect identify $\tilde{H}$ and $H$:
\be
\langle s_{\theta_i}(w) \tilde{s}_{\theta_i} (\bar{w}) \rangle = (w - \bar{w})^{-\theta_i^2}.
\ee
Using bosonisation the calculation of correlation functions between worldsheet spinors and twist fields is straightforward and essentially identical to the untwisted case:
\be
\langle \prod_i e^{i q_i \cdot H(z_i)} \rangle = \prod_{i<j} z_{ij}^{q_i \cdot q_j}, \hbox{ for } \sum_i q_i =0.
\ee

\subsubsection*{Twisted vertex operator}

To calculate CFT correlation functions involving blow-up modes that resolve
orbifold singularities we need the vertex operator that resolves the blow-up.
This is the closed string vertex operator for a twisted scalar $V_{\textrm{tw}}^{(-1, -1)}(w, \wb)$ and combines the fermionic and the bosonic
twists. The appropriate vertex operator of correct conformal weights $h=1, \tilde{h}=1$ is
\begin{equation}
V_{\textrm{tw}}^{(-1, -1)}(w, \wb) = \ e^{- \phi (w)} e^{- \tilde{\phi}(\wb)} \prod_{i=1}^3 s_{\theta_i}(w) \tilde{s}_{\theta_i}(\wb) \tw(w, \wb) \ e^{i k \cdot X(w, \wb)} \ .
\end{equation}
The worldsheet supercurrents $T_F(z)$ and $\tilde{T}_F(\bar{z})$ are single-valued on transport around $V_{\textrm{tw}}^{(-1, -1)}(w, \wb)$.
 The conformal weights are given by the sum of the ghost, fermionic and bosonic weights: $h=\tilde{h}= \half + \sum_{i=1}^3 \left(\frac{1}{2} \theta_i^2 +\frac{1}{2} \theta_i (1- \theta_i) \right)=\frac{1}{2} \sum_{i=1}^3 \theta_i$. In a consistent orbifold theory that preserves $\mathcal{N}=1$ supersymmetry the sum over twists obeys $\sum_{i=1}^3 \theta_i=0 \textrm{ mod } 1$, and so we can choose the twists to satisfy $\sum_{i=1}^3 \theta_i=1$.

On the disk we can factorize the above operator into two factors that only contain left-moving or right-moving fields by splitting $\tw(w, \wb) =\tw(w) \tilde{\sigma}_{\theta_i}(\wb)$. Further, as the momentum $k$ is purely in the external directions and hence parallel to the D-branes we can write the momentum exponential as $e^{i k \cdot X(w, \wb)}=e^{i k \cdot X(w)} e^{i k \cdot \tilde{X}(\wb)}$.

\subsection{Kinematics}

In this paper we evaluate amplitudes involving one closed string and either two or three open strings.
From a kinematic point of view one closed string behaves like two open strings and thus, naively, we expect our results to exhibit the kinematical structure of four- and five-point functions of open strings. We will only consider massless string states in our calculations and thus the inner product of any momentum with itself vanishes.

We first examine scattering between two open strings with momenta $p_1$ and $p_2$ and a closed string of momentum $q$.
All momenta lie parallel to the brane. The exponents of open string vertex operators involve $k_1 = 2 p_1$ and $k_2 = 2p_2$
(as both holomorphic and antiholomorphic parts contribute). It is useful to view the closed string vertex operator as a left-moving
vertex operator located at $w$ and a right-moving vertex operator at $\bar{w}$, with momenta $k_3 = q$ and $k_4 = q$, in effect turning the amplitude into a 4-point amplitude. We can write the equation of momentum conservation as
\begin{equation}
k_1+k_2 + 2q=0 \ .	
\end{equation}
We can conveniently define Mandelstam variables as for a 4-point function:
\begin{equation}
	s=k_1k_2, \qquad t=k_1k_3, \qquad u=k_1k_4 \ .
\end{equation}
However these kinematic variables are not independent.
Momentum conservation leads to $s+t+u=0$ and furthermore as $k_3=k_4=q$, we also have
\begin{equation}
	t=u, \qquad s=-2t,
\end{equation}
leaving only one independent kinematic variable. Note that for on-shell zero mass particles, kinematics imply $s=t=u=0$. One can interpret the
three point function by allowing the momenta to go off-shell, but this is slightly ambiguous as if $k^2 \neq 0$ the conformal weight of the operator
is incorrect. The only fully clean way to resolve this is go to 4-point functions where no such ambiguity exists.

We will also consider the 4-pt amplitude involving three open strings (momenta $p_1, p_2, p_3$) and one closed string (momentum $q$). Again, we will only consider massless string states such that $k_i^2=q^2=0$. Momentum conservation now requires that ($k_i = 2p_i$)
\begin{equation}
k_1+k_2+k_3 + 2q=0 \ .	
\end{equation}
Mandelstam variables are given by
\begin{equation}
	s=k_1k_2, \qquad t=k_1k_3, \qquad u=2k_1q \ .
\end{equation}

\subsection{Classical Contributions to Correlation Functions}

There are two parts to worldsheet correlation functions, classical and quantum.
In several twist field correlators an important element is the calculation of the classical contribution to the
correlation function. The classical contribution arises
from worldsheet embeddings that satisfy the classical equations of motion,
\be
\partial \bar{\partial} X = 0,
\ee
and also satisfy the appropriate boundary conditions. Its magnitude is given by $e^{-S_{cl}}$, where the classical action is
\be
S_{cl} = \frac{1}{2 \pi \alpha'} \int d^2 z \left( \partial X_{cl} \bar{\partial} \bar{X}_{cl}
+ \bar{\partial} X_{cl} \partial \bar{X}_{cl} \right) \ .
\ee
Here we revert to denoting the complexified bosonic coordinates by $X$ instead of $Z$ to avoid confusion with worldsheet coordinates in this section.

We are considering correlators between open strings attached to a D3-brane and bulk twist fields.
The quantum correlator arises from fluctuations about the classical solution. Twist correlators require summing
over all classical solutions and weighting by the action of each solution.

A D3-brane located at $X_{i} = x_{i}$ imposes the disk boundary condition
\be
X_i(\hbox{Im}(z) = 0) = x_i.
\ee
In space-time any given twist field is associated to an orbifold singularity located at
$(x^{tw}_1, x^{tw}_2, x^{tw}_3)$. Inserting this twist field at the world sheet location $z_0$, $\zb_0$
implies that the worldsheet embedding $X(z, \zb)$ should satisfy the following conditions:
\bea
\label{yoda}
X_i(z_0, \zb_0) & = & (x^{tw}_1, x^{tw}_2, x^{tw}_3), \nn \\
\partial_z X \sigma_{\theta}(z_0, \zb_0) & \sim & (z-z_0)^{-1+\theta} (z-\zb_0)^{-\theta}, \nn \\
\partial_z \bar{X} \sigma_{\theta}(z_0, \zb_0) & \sim & (z - z_0)^{-\theta} (z - \zb_0)^{-1+\theta},\nn \\
\partial_{\bar{z}} X \sigma_{\theta}(z_0, \zb_0) & \sim & (\bar{z} - z_0)^{-1+\theta} (\bar{z} - \zb_0)^{-\theta},\nn \\
\partial_{\bar{z}} \bar{X} \sigma_{\theta}(z_0, \zb_0) & \sim & (\zb-z_0)^{-\theta} (\zb-\zb_0)^{-1+\theta}
\eea
which follow from the singular behaviour of the bosonic fields on the disk (\ref{OPED} - \ref{OPED3}).

For the general case where the twist field is located away from the D3-brane stack,
it is difficult to obtain the classical solutions. However, in the case that the twist field corresponds to the same singularity
where the D3-brane stack is located (i.e. $x_i^{tw} = x_i$) a trivial classical solution with vanishing classical action exists,
$X_{cl}^i(z, \zb) = x^i$ with $\partial X_{cl} = 0$. There may be other classical solutions, but these will be exponentially suppressed
as $e^{-2 \pi R^2}$ and can be neglected.

For the quantum calculations in sections 3 and 4 below we will implicitly assume the trivial classical solution.
In section 5 we will reconsider the case where the twist field
is geometrically separated from the D3-brane stack, when the classical solutions play an important role.

\subsection{Twist Fields and the Worldsheet Boundary}
\label{branchpt}

One of the more subtle consequences of the presence of a twist field on the disk manifests itself on the boundary. The insertion of a single twisted vertex operator in the bulk of the disk introduces a branch cut into the correlation function. The branch cut runs
across the disk worldsheet with its other end on the disk boundary. Even though the overall result for the amplitude must be independent of the position of the branch point, we have to choose a specific boundary locus for it for explicit calculations. It is most practical to choose the branch cut to touch the disk boundary at infinity. This choice of infinity is equivalent
to using the canonical definition of $(z-z_0)^{\theta}$, with the branch cut along the negative real axis, in eq. (\ref{ZZtw}).
We then have the basic tools to calculate the CFT correlation function with the twist field inserted.
The presence of a branch point on the boundary however introduces an extra subtlety which we now discuss.

Let us first recall the prescription for calculating correlation functions on the disk without a twist field insertion.
We are required to sum over all cyclically inequivalent orderings of the vertex operators on the boundary together with
appropriate traces over Chan-Paton factors with the same ordering.
We can fix a certain number of operators depending on the worldsheet reparametrisation invariance: on the disk this is given by SL$(2, \mathbb{R})$ and
allows us to fix three real values for the positions of disk vertex operators.
As cyclically equivalent orderings are identical, for three boundary operators $A$, $B$, $C$ there are effectively only two orderings to be considered
$A_0 B_1 C_{\infty}$ and $B_0 A_1 C_{\infty}$.

\begin{figure}
\begin{center}
\includegraphics[height=5cm]{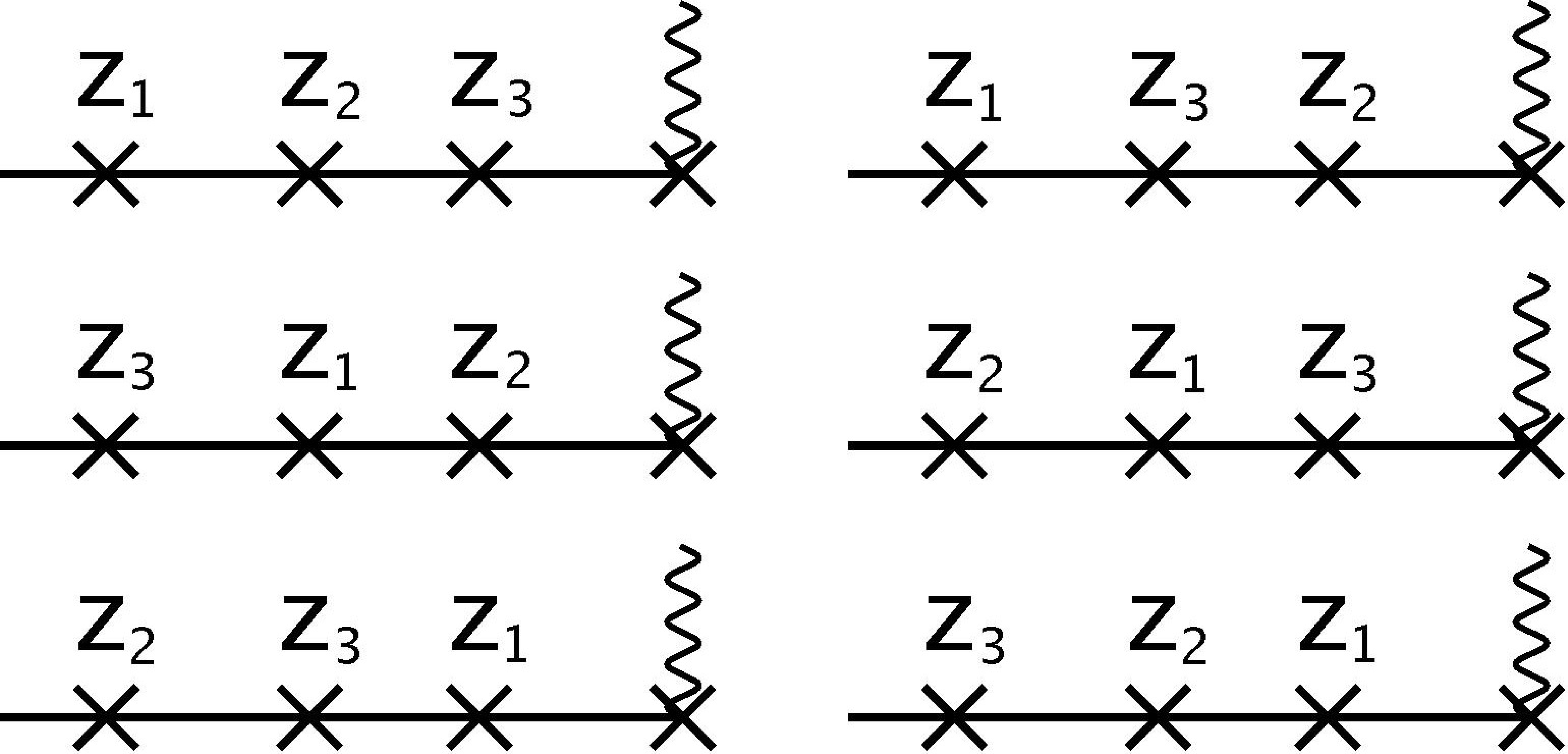}
\caption{The six possible (and distinct) orderings of three vertex operator in relation to the location of the branch point
on the boundary of the disk. The branch point is fixed to be at $\infty$.}
\label{allorder}\end{center}\end{figure}
This situation is to be contrasted with our case where there is also a branch point on the boundary. We follow a suggestion in the appendix of \cite{Douglas:1996sw}. The branch point has to be considered effectively as an additional operator when considering orderings.
We now need to sum over all cyclically inequivalent orderings of both the operators and the branch point.
Our analysis for three operators on the boundary is modified when a twist is inserted: there are now six orderings of the four objects on the boundary which are shown in figure \ref{allorder}, all of which need to be evaluated.
 Figure \ref{allorder} applies for our calculation of a Yukawa coupling with blow-up mode insertion. We we will discuss the details of
 fixing the SL$(2, \mathbb{R})$-invariance in section 4.

Another situation that occurs in our calculations involves two open string vertex operators and one branch point on the boundary.
 In this case there are two quantum correlators associated to the two different cyclic orderings of the vertex operators and the branch point.

Once we have calculated individual correlators corresponding to each ordering we need to combine them to obtain the final result.
A vertex operators moved through a branch point acquires a phase and we must account for this phase when summing the individual amplitudes.
Different orderings correspond to moving a vertex operator through the branch point and so we sum each correlator weighted by a compensating
phase.

These phases can be understood by considering the effect of the twist on the Chan-Paton factor. An orbifold twist does not only act geometrically, it also acts on the Chan-Paton factors via a twist matrix $\gamma_\theta$. The presence of the twist operator leads to an effective insertion of the
twist matrix $\gamma_\theta$ on the boundary to cancel the phase associated to the branch point.
Differently orderings of vertex operators correspond to different Chan-Paton traces, e.g. $\hbox{Tr}(T_1 T_2 T_3 \gamma_{\theta})$
as opposed to $\hbox{Tr}(T_3 T_1 T_2 \gamma_{\theta})$.  Orderings are related by commutation relations
for the Chan-Paton factor $T_i$ and the twist matrix $\gamma_\theta$, which are easily found by considering
the orbifold spectrum. Thus these operators commute up to a phase:
\be
\label{CPcommutator}
T_i \gamma_{\theta} = e^{-2 \pi i \theta_i} \gamma_{\theta} T_i \quad \quad \textrm{for bosons.}
\ee
It is this phase that needs to be included into our sum over the amplitude fragments.

One can alternatively derive these phases by examining the behaviour of vertex operators as they are moved around the disk boundary. Due to the monodromy introduced by the twist field, the quantum amplitude obtains a phase factor $e^{-2 \pi i \theta}$ when a vertex operator is
moved once around the disk. As the amplitude has to be single-valued, we can account for this through a relative phase offset when a vertex operator
moves through the branch point.

We will employ these procedures in the calculations that follow in section 3 and section 4 below.

\section{Preliminary Calculations: Two Open Strings and One Twist Field}

 All amplitudes factor into a quantum correlator and a contribution from the classical action.
 In sections 3 and 4 we focus on the quantum correlators and present results for calculations including two or three matter fields and one blow-up mode.

 The matter lives on a stack of D3-branes at an orbifold singularity in the compact space. The blow-up mode can be located either
 at the same orbifold singularity as the matter or alternatively one geometrically distant. The quantum amplitude is a CFT calculation on the disk worldsheet and is thus ignorant of the embedding in spacetime, and so valid independent of the blow-up mode location. In the case that matter and the blow-up mode share the same orbifold fixed point, the contribution from the classical action is trivial and the quantum correlator gives the full result. If the stack of D3-branes is distant in spacetime from the blow-up mode, the classical action is more complicated. For now, we proceed with presenting the quantum results in chapters 3 and 4 and will discuss the classical part for distant blow-up modes in section 5.

\subsection{Gauge Kinetic Term in the presence of a Twist}

In this section we examine the dependence of the gauge kinetic function on twisted K\"ahler moduli. The amplitude $\langle A A \ b \rangle$ we calculate involves two gauge fields on the boundary and one twisted closed string in the disk interior.

This calculation does not fall into the main narrative of sequestering in local models, but is useful for providing a test and simple application
of the technology we use.

To obtain a total ghost charge of $-2$ as required on the disk we choose vertex operators for the gauge bosons in the zero ghost picture. The appropriate vertex operators are then:
\begin{align}
V_{A^{a_1}}^{(0)}(z_1) =& \ \lambda^{a_1} {\xi_1}_\mu \left[\pd X^\mu + i (k_1 \cdot \psi) \psi^\mu) \right] e^{i k_1 \cdot X}(z_1), \\
V_{A^{a_2}}^{(0)}(z_2) =& \ \lambda^{a_2} {\xi_2}_\nu \left[\pd X^\nu + i (k_2 \cdot \psi) \psi^\nu) \right] e^{i k_2 \cdot X}(z_2), \\
 \nn V_{\textrm{tw}}^{(-1, -1)}(w, \wb) =& \ e^{- \phi (w)} e^{- \tilde{\phi}(\wb)} \displaystyle\prod_{j=1}^3 \tw(w, \wb) \\& \  e^{- i q_3 \cdot H(w)} e^{- i \tilde{q}_3 \cdot \tilde{H}(\wb)} \ e^{- i k_3 \cdot X(w)} e^{- i k_3 \cdot X(\wb)} \ .
\end{align}
Chan-Paton factors are denoted by $\lambda^{a}$ and the polarisation vectors ${\xi}_\mu$ obey ${\xi}_\mu{k}^\mu=0$. The H-charges appearing in the vertex operator for the twisted string are $q_3=(\theta_1,\theta_2,\theta_3)$ and $\tilde{q}_3=(-\theta_1,-\theta_2,-\theta_3)$. We ignore the Chan-Paton factors in the following worldsheet computation as they will only provide a constant factor.

We begin the calculation by observing that the vertex operators for the gauge bosons only consist of fields in external directions and, further, the only internal fields present are the twist fields themselves. Thus the twist fields will only contract with themselves and leave the other
correlators unaffected:
$$
\la \prod_i s_{\theta_i}(w) \tilde{s}_{\theta_i}(\wb) \ra \la \prod_i \tw(w, \wb)\ra= {(w-\wb)}^{-1}.
$$
 As a result all dependence on the twists already vanishes at this stage.

Continuing, since the vertex operators for the gauge fields are a sum of two parts, the full amplitude $\mathcal{A}$ consists of four terms, two of which do not contribute: the cross-contractions between the terms involving $\pd X$ and the terms including $\psi$ will vanish due to $\langle :(k \psi) \psi^\mu : \rangle=0$. We are left with two amplitudes to evaluate: one involving the worldsheet spinors
$$
\la (k_1 \cdot \psi) \psi^\mu (z_1)  (k_2 \cdot \psi) \psi^\nu (z_2) e^{ik_1 \cdot X (z_1)} e^{i k_2 \cdot X (z_2)} e^{-i k_3 \cdot X (w)}
e^{-i k_4 \cdot X (\bar{w})} \ra,
$$
whereas the other term involves the bosonic correlator
$$
\la \pd X^\mu e^{i k_1 \cdot X}(z_1) \pd X^\nu e^{i k_2 \cdot X}(z_2) e^{- i k_3 \cdot X}(z_3) e^{- i k_4 \cdot X}(\zb_3) \ra.
$$

The correlators can be evaluated using the basic correlators collected in section \ref{CFTBasics}. The $SL(2, \mathbb{R})$ invariance on the disk
allows us to fix three worldsheet coordinates. This can be done straightforwardly.
The complications of section \ref{branchpt}, requiring the presence of a branch cut, do not apply. As the vertex operators on the boundary only contain fields in the external directions while the twist only acts internally the amplitude is not branched. There is then
no branch point on the boundary and we only need consider one ordering of the vertex operators. A useful choice for the $SL(2, \mbb{R})$ fixing
 is \cite{Hashimoto:1996bf, 0404134}:
\begin{equation}
z_1=x \quad , \quad z_2=-x, \quad , \quad w=i \quad , \quad \wb=-i,
\end{equation}
which leads to a c-ghost contribution of $2i(x^2+1)$. The
bosonic momentum exponentials also provide a term ${\left(\frac{x^2+1}{4x} \right)}^{2t}$. The full result for the amplitude is
\begin{align}
\mathcal{A} \propto \ & -i 2^{-1-4t} \xi_1 \xi_2 (1+2t) \frac{1}{4} \int_{0}^{\infty} \textrm{d}x \frac{{(x-i)}^{2t+1}{(x+i)}^{2t+1}}{x^{2t+2}}
\nonumber \\
& -i 2^{-1-4t} (\xi_1 k_2)(\xi_2 k_1) \int_{0}^{\infty} \textrm{d}x \frac{{(x-i)}^{2t-1}{(x+i)}^{2t-1}}{x^{2t}} \ .
\end{align}
The first term originates from the worldsheet spinor correlator and the contraction between $\pd X^\mu $ and $\pd X^\nu$. The second term comes purely from contractions of $\pd X^\mu$ and $\pd X^\nu$ with the momentum exponentials. The various terms have been simplified using $-2\xi_1 k_3=\xi_1k_2$ and $-2\xi_2 k_3=\xi_2 k_1$ (the term proportional to $(\xi_1 k_2)(\xi_2 k_1)$ coming from the spinor correlator is cancelled by a contribution from $\la \pd X^\mu \pd X^\nu \ldots \ra$).

These integrals can be evaluated using the following expressions
\begin{align}
\label{GammaInt}
\nn I(\delta, \alpha) &= \int_0^{\infty} \textrm{d}x \ x^{\delta-1} {(x-i)}^{\alpha - \delta} {(x+i)}^{- \alpha - \delta} \\
& = \sqrt{\pi} 2^{-\delta} e^{-\frac{1}{2} \pi i \alpha} \frac{\Gamma \left(\frac{\delta}{2} \right)\Gamma \left(\frac{1}{2}+\frac{\delta}{2} \right)}{\Gamma \left(\frac{1}{2}+\frac{\delta}{2}-\frac{\alpha}{2} \right)\Gamma \left(\frac{1}{2}+\frac{\delta}{2}+\frac{\alpha}{2} \right)} \ .
\end{align}
We then obtain
\begin{equation}
\mathcal{A} \propto i \pi \left[\xi_1 \xi_2 t + \frac{1}{2}(\xi_1 k_2)(\xi_2 k_1) \right] t \frac{\Gamma(-2t)}{{\Gamma(1-t)}^2} \ .
\end{equation}
The gamma functions can be expanded for small momenta to yield $t \frac{\Gamma(-2t)}{{\Gamma(1-t)}^2}= - \frac{1}{2} + \mathcal{O}(k^2)$. Rewriting the open string momenta in terms of the physical momenta $p_1=k_1/2$, $p_2=k_2 /2$ and recalling that $-2t=s$ we are able to extract the kinematical factor for gauge bosons:
\begin{equation}
\mathcal{A} \propto i 2 \pi \left[\half (p_1 p_2) (\xi_1 \xi_2) - \half (p_1 \xi_2)(p_2 \xi_1) \right] +\mathcal{O}(p^4) \ .
\end{equation}
Blow-up moduli are known to enter the gauge kinetic function for branes at a singularity, and so this result is consistent with expectations.
This provides a useful check on the formalism and a confirmation of the disk single-twist correlator.

\subsection{Matter Metric in the presence of a Twist}
\label{mattermetric}

In this section we examine how blow-up modes enter the matter metric for charged matter fields $C_i$ and $\bar{C}_i$. To this end we calculate the disk amplitude $\langle C_i \bar{C}_i b \rangle$ with the two matter fields inserted at the boundary and the twisted closed string in the bulk.

To get a total ghost charge of $-2$ on the disk we take the twisted closed string in the $(-1,-1)$ ghost picture and choose vertex operators with zero ghost charge for the matter fields:
\begin{align}
V_{C_i}^{(0)}(z_1) =& \ \lambda^{\hphantom{\dagger}} \left[\pd Z_i + i (k_1 \cdot \psi) \Psi_i) \right] e^{i k_1 \cdot X}(z_1), \\
V_{\bar{C}_i}^{(0)}(z_2) =& \ \lambda^{\dagger} \left[\pd \Zb_i + i (k_2 \cdot \psi) \bar{\Psi}_i) \right] e^{i k_2 \cdot X}(z_2), \\
 \nn V_{\textrm{tw}}^{(-1, -1)}(z_3, \zb_3) =& \ \gamma_\theta \ e^{- \phi (z_3)} e^{- \tilde{\phi}(\zb_3)} \displaystyle\prod_{j=1}^3 \tw(z_3, \zb_3) \\& \  e^{i q_3 \cdot H(z_3)} e^{i q_4 \cdot \tilde{H}(\zb_3)} \ e^{i k_3 \cdot X(z_3)} e^{i k_4 \cdot X(\zb_3)} \ .
\end{align}
The H-charges appearing are $q_3=(\theta_1, \theta_2, \theta_3)$ and $q_4=(-\theta_1, -\theta_2, -\theta_3)$. We ignore the Chan-Paton factors at first and focus on the CFT calculation. We get the following contributions to this amplitude: we define $\mc{A}_1$ as the result of contracting the $\partial Z$ terms of both matter field vertex operators, while contracting the second terms will lead to the expression $\mc{A}_2$. Cross-correlations will vanish due to normal ordering leaving only these two terms.

We begin our analysis by studying the amplitude $\mc{A}_1$. It consists of the following disk correlation functions:
\begin{align}
\nn \mc{A}_1 \propto & \ \la e^{- \phi(z_3)} e^{-\tilde{\phi}(\zb_3)} \ra \la e^{i q_3 \cdot H(z_3)} e^{i q_4 \cdot \tilde{H}(\zb_3)} \ra \la e^{i k_1 \cdot X(z_1)} e^{i k_2 \cdot X(z_2)} e^{i k_3 \cdot X(z_3)} e^{i k_4 \cdot X(\zb_3)} \ra \\
& \ \la \pd Z_i (z_1) \pd \Zb_i (z_2) \displaystyle\prod_{j=1}^3 \tw(z_3, \zb_3) \ra \ .
\end{align}
The correlators appearing in the first row can be evaluated using the basic techniques of section \ref{CFTBasics}. The last factor is the correlation function between the bosonic coordinate and twist fields which we evaluated before in (\ref{ZZtw}) and which contributes two terms: we define $\mc{A}_{1b}$ as the part that is directly proportional to $\theta_i$ whereas $\mc{A}_{1a}$ is the remaining piece.

We now turn to the amplitude $\mc{A}_2$ arising from the contraction of the second terms of the matter vertices before combing all our results. In terms of the individual worldsheet correlators we obtain:
\begin{align}
\nn \mc{A}_2 \propto & \ - \la e^{- \phi(z_3)} e^{-\tilde{\phi}(\zb_3)} \ra \la (k_1 \psi) \psi^\mu (z_1) (k_2 \psi) \psi^\nu (z_2) \ra \la e^{i k_1 \cdot X(z_1)} e^{i k_2 \cdot X(z_2)} e^{i k_3 \cdot X(z_3)} e^{i k_4 \cdot X(\zb_3)} \ra \\
& \ \la e^{i q_1 \cdot H(z_1)} e^{i q_2 \cdot H(z_2)} e^{i q_3 \cdot H(z_3)} e^{i q_4 \cdot \tilde{H}(\zb_3)} \ra \la \displaystyle\prod_{j=1}^3 \tw(z_3, \zb_3) \ra, \
\end{align}
where we bosonised the internal spinors $\Psi_i(z_1)$ and $\bar{\Psi}_i(z_2)$ with H-charges $q_1=(\delta_{1i}, \delta_{2i}, \delta_{3i})$ and $q_2=(-\delta_{1i},-\delta_{2i},-\delta_{3i})$. The bosonic twists only contribute their self-correlator (\ref{self}) and
the remainder is straightforward. We find that $\mc{A}_2$ is proportional to the same worldsheet integral as $\mc{A}_{1a}$.

We next use the $SL(2, \mathbb{R})$ invariance to fix three real parameters amongst the positions of the vertex operators.
There is now a branch point in the amplitude and
following section \ref{branchpt} to capture all contributions to this amplitude we need to compute two correlators with different orderings of the
vertex operators. In order to include all orderings we fix the worldsheet coordinates as follows \cite{Hashimoto:1996bf, 0404134}:
\begin{equation}
\label{fixHashimoto}
z_1=x \quad , \quad z_2=-x, \quad , \quad z_3=i \quad , \quad \zb_3=-i.
\end{equation}
Then $x > 0$ corresponds to the first ordering and $x < 0$ corresponds to the second ordering.
In both cases we need to include the c-ghost contribution $\la c(z_2) c(z_3) \tilde{c}(\zb_3) \ra = (z_2-z_3)(z_2-\zb_3)(z_3-\zb_3) = 2i(x^2+1)$.

\subsubsection*{First way of ordering}
 With the first way of ordering the resulting expressions are:
\begin{align}
&\mc{A}_{1a}+\mc{A}_2 & \propto & \ (1+2t) \ \frac{i}{4} \ 2^{-1-4t} \ &&e^{2 \pi i \theta_i} \ \int_0^{\infty} \textrm{d}x \frac{{(x+i)}^{2t+1-2\theta_i}{(x-i)}^{2t +1+2\theta_i}}{x^{2t+2}}, \\
&\mc{A}_{1b} & \propto & \ \theta_i \ 2^{-1-4t} \ &&e^{2 \pi i \theta_i} \ \int_0^{\infty} \textrm{d}x \frac{{(x+i)}^{2t+1-2\theta_i}{(x-i)}^{2t-1+2\theta_i}}{x^{2t+1}} \ .
\end{align}
These can be evaluated using eq. (\ref{GammaInt}), in terms of which we obtain
\begin{align}
\mc{A}_{z_1=x} & \propto e^{2 \pi i \theta_i} \ 2^{-1-4t} \left( (1+2t) \ \frac{i}{4} \ I(-1-2t, 2\theta_i) + \theta_i \ I(-2t, 2\theta_i -1) \right) \\
& \propto i \pi e^{\pi i \theta_i} \Gamma\left(-2t \right) \left[ \frac{\Gamma(1-\theta_i-t) + \theta_i \ \Gamma(-\theta_i-t)}{\Gamma(-\theta_i-t)\Gamma(\theta_i-t)\Gamma(1-\theta_i-t)} \right] \ .
\end{align}
We expand in powers of momentum to obtain
\begin{equation}
\mc{A}_{z_1=x} \propto \frac{i}{2} e^{\pi i \theta_i} \sin (\pi \theta_i) \left(1 + t \left[2 \gamma_E + \psi(\theta_i) + \psi(1-\theta_i) \right] + \mc{O}(t^2)\right),
\end{equation}
where $\psi(z) = \frac{\textrm{d}}{\textrm{d}x} \Gamma(z)$. This concludes the CFT calculation of this part of the amplitude.

\subsubsection*{Second way of ordering}
We now repeat this analysis for the second way of ordering the vertex operators with respect to the branch point. We fix worldsheet coordinates as in \eqref{fixHashimoto}, but now we integrate over $x < 0$. Performing the calculation with this choice of worldsheet coordinates we obtain:
\begin{align}
\mc{A}_{z_1=-x} & \propto e^{-2 \pi i \theta_i} \ 2^{-1-4t} \left( (1+2t) \ \frac{i}{4} \ I(-1-2t, -2\theta_i) - \theta_i \ I(-2t, -2\theta_i +1) \right) \\
& \propto i \pi e^{-\pi i \theta_i} \Gamma\left(-2t \right) \left[ \frac{\Gamma(1-\theta_i-t) + \theta_i \ \Gamma(-\theta_i-t)}{\Gamma(-\theta_i-t)\Gamma(\theta_i-t)\Gamma(1-\theta_i-t)} \right] \ .
\end{align}
Having arrived at this expression the only difference to the result for the other ordering is in the overall phase factor, whereas the gamma functions are identical. Hence the expansion for small momentum is identical apart from the phase factor:
\begin{equation}
\mc{A}_{z_1=-x} \propto \frac{i}{2} e^{-\pi i \theta_i} \sin (\pi \theta_i) \left(1 + t \left[2 \gamma_E + \psi(\theta_i) + \psi(1-\theta_i) \right] + \mc{O}(t^2)\right) \ .
\end{equation}

\subsubsection*{Combination of results}

When combing the results of the two previous sections we have to account for the fact that the vertex operators on the boundary are ordered differently with respect to the branch point introduced by the twist as laid out in section \ref{branchpt}. Here, we can reinstate the traces over the Chan-Paton factors to do this for us: the first way of fixing the vertex operators on the boundary provides a trace $\textrm{Tr}(\lambda^\dagger \lambda \gamma_\theta)$ whereas the second analysis contains the factor $\textrm{Tr}(\lambda \lambda^\dagger \gamma_\theta)$. We can write both our results using one common trace factor by noting that
\begin{equation}
\textrm{Tr}(\lambda \lambda^{\dagger} \gamma_{\theta}) = e^{2 \pi i \theta_i} \textrm{Tr}(\lambda^{\dagger} \lambda \gamma_{\theta}) \ .
\end{equation}
Here we used the commutator between the Chan-Paton factor for a boson and the orbifold twist $\gamma_\theta$ \eqref{CPcommutator}. We can now
 combine our partial expressions for the quantum correlation function to arrive at the final result:
\begin{equation}
\label{CCb}
\mc{A}_{\textrm{full}} \propto \ \textrm{Tr}(\lambda^{\dagger} \lambda \gamma_{\theta}) \ i e^{\pi i \theta_i} \sin (\pi \theta_i) \left(1 + t \left[2 \gamma_E + \psi(\theta_i) + \psi(1-\theta_i) \right] + \mc{O}(t^2)\right) \ .
\end{equation}
As a side comment we note that the result is identical to the outcome of a calculation of the dependence of the twisted matter-metric on untwisted K\"ahler moduli \cite{0404134}.

 Although a warm-up, this calculation has required use of the full set of calculational tools necessary for working with twist fields on the disk.  Our expression will furthermore be useful in what follows: we will use the above answer as a consistency-check on a lower-point limit of a Yukawa coupling with a twist insertion.

\section{Yukawa Couplings in the presence of a Twist}

We now address our main topic, the calculation of corrections to the Yukawa couplings. We calculate a string scattering amplitude on the disk with open strings on the boundary and an insertion of a blow-up mode in the bulk. We will present the full quantum result for this amplitude including the trace over Chan-Paton factors.

\subsection{Quantum correlator}
The quantum correlator of a Yukawa coupling in the presence of a twist will involve two fermionic and one bosonic vertex operators on the boundary of the disk and one twist vertex operator inserted in the bulk. In the canonical picture these vertex operators are:
\bea
\label{vo1}
V_{\psi_1}^{(-\frac{1}{2})}(z_1)  & = &  \lambda_1 \ e^{-\frac{1}{2} \phi (z_1)} S^{\pm}(z_1) e^{i q_1 \cdot H}(z_1) e^{i k_1 \cdot X}(z_1), \\
V_{\psi_2}^{(-\frac{1}{2})}(z_2) & = &  \lambda_2 \ e^{-\frac{1}{2} \phi (z_2)} S^{\mp}(z_2) e^{i q_2 \cdot H}(z_2) e^{i k_2 \cdot X}(z_2), \\
V_{\phi}^{(-1)}(z_3) & = & \lambda_3  e^{- \phi (z_3)} e^{i q_3 \cdot H}(z_3) e^{i k_3 \cdot X}(z_3), \\
V_{\textrm{tw}}^{(-1, -1)}(w, \bar{w}) & = & \gamma_\theta \ e^{- \phi (w)} e^{- \tilde{\phi}(\bar{w})} \displaystyle\prod_{j=1}^3 \tw(w, \bar{w}) \nn \\
 & &  \ti e^{i q_4 \cdot H(w)} e^{i q_5 \cdot \tilde{H}(\bar{w})} \ e^{i k_4 \cdot X(w, \bar{w})},
 \label{vo4}
\eea
where $\lambda_i$ denote Chan-Paton factors which we will ignore in this part of the calculation. The external spinors $S^{\pm}$ are given by $e^{\pm \frac{i}{2} (H_1+H_2)}$ and the internal H-charges are $q_1=(\frac{1}{2},-\frac{1}{2},-\frac{1}{2})$, $q_2=(-\frac{1}{2},\frac{1}{2},-\frac{1}{2})$, $q_3=(0,0,1)$, $q_4=(\theta_1,\theta_2,\theta_3)$ and $q_5=(-\theta_1,-\theta_2,-\theta_3)$ leading to the following overall H-charge structure:
\begin{align}
V_{\psi_1}^{(-\frac{1}{2})}(z_1) \ \sim & \ |++\rangle \otimes  |+--\rangle, \\
 V_{\psi_2}^{(-\frac{1}{2})}(z_2) \ \sim & \ |--\rangle \otimes  |-+-\rangle, \\
 V_{\phi}^{(-1)}(z_3) \ \sim & \ |0 \, 0\rangle \otimes  |0 \, 0 (++) \rangle, \\
 V_{\textrm{tw}}^{(-1)}(w) \ \sim & \ |0 \, 0\rangle \otimes  |\theta_1, \ \theta_2, \ \theta_3 \rangle, \\
 V_{\textrm{tw}}^{(-1)}(\bar{w}) \ \sim & \ |0 \, 0\rangle \otimes  |- \theta_1, \ -\theta_2, \ -\theta_3 \rangle \ .
\end{align}
Lorentz invariance allows us to impose without loss of generality $S^1 = |++\rangle$ and $S^2 = |--\rangle$. This restriction is equivalent
to boosting to a frame with $k_1 = (k, k, 0, 0)$ and $k_2 = (k, -k, 0, 0)$. To see this, note that the physical state condition
$(k \cdot \Gamma) | \psi \rangle = 0$ gives
$$
(k \cdot \Gamma) | \psi \rangle = (k_0 \Gamma^0 \pm k_1 \Gamma^1) | \psi \rangle = -k_1 \Gamma^0(\Gamma^0 \Gamma^1 \mp 1) | \psi \rangle
= - 2k_1 \Gamma^0 (S_0 \mp 1/2) | \psi \rangle = 0,
$$
and so $S_0 = \pm 1/2$ for $k_1 = \pm k_0$. The GSO conditions then imply $S^1 = |++\rangle$ and $S^2 = |--\rangle$.
We also note that the physical state conditions imply $k_1^{1+} = k_1^{2+} = k_1^{2-} = 0$
and $k_2^{1-} = k_2^{2-} = k_2^{2+} = 0$.

However, the vertex operators (\ref{vo1}) to (\ref{vo4}) do not possess the
correct overall ghost charge (-2) for a disk correlation function.
To obtain a ghost charge of $-2$ we choose to picture-change two vertex operators on the boundary of the disk.
An alternative would be to picture-change the bulk twist operator. However this would be
technically more involved and require correlators involving excited twist fields.
 We will picture change the bosonic vertex operator $V_{\phi}(z_3)$ and the second fermionic vertex operator $V_{\psi_2}(z_2)$,
 modifying their ghost charges as $-1 \rightarrow 0$ and $-\frac{1}{2}\rightarrow \frac{1}{2}$ respectively.

Picture-changing of a vertex operator on the boundary is performed by evaluating the limit $\lim_{z \rightarrow w} e^{\phi(z)} T_{F}(z) V^{(c)}(w)$ where divergent terms of $\mathcal{O}{(z-w)}^{-1}$ are dropped. The picture-changing operator involves the worldsheet supercurrent $T_F$ which takes the following form on the boundary of the disk:
\begin{equation}
T_F(z) = \pd X_{\mu}\psi^{\mu}(z) + \sum_{i=1}^{3} \left[\pd \Zb_i \Psi_i (z) + \pd Z_i \bar{\Psi}_i(z) \right].
\end{equation}
The picture-changed amplitude obtains contributions from picture-changing either in the internal or external directions.
Before describing these, we first discuss generalities that apply in both cases.

\subsection{Generalities}

There is a external momentum correlator, given by
\be
\langle e^{i k_1 \cdot X(z_1)} e^{i k_2 \cdot X(z_2)} e^{i k_3 \cdot X(z_3)} e^{i k_4 \cdot X(w, \bar{w})} \rangle.
\ee
As we are studying amplitudes involving D3-branes all momenta are along the external directions.
We denote $k_1 \cdot k_2 = s$, $k_1 \cdot k_3 = t$, $k_1 \cdot k_4 = u$. Momentum conservation
requires
\be
k_1 + k_2 + k_3 +k_4 = 0.
\ee
As we are studying a 4-point amplitude, we can without ambiguity work with all momenta on-shell
$(k_1^2 = k_2^2 = k_3^2 = k_4^2 = 0)$ and at finite $s, t$ and $u$, subject to $s + t + u = 0$.\footnote{
For 2-point or 3-point amplitudes it is necessary to use an off-shell prescription, and
work at finite $k_i \cdot k_j$ before taking the limit $k_i \cdot k_j \to 0$. For 4- and higher point amplitudes
there is no need to do this and the results are unambiguous.} We then have
\be
\langle e^{i k_1 \cdot X(z_1)} e^{i k_2 \cdot X(z_2)} e^{i k_3 \cdot X(z_3)} e^{i k_4 \cdot X(w, \bar{w})} \rangle
= \vert z_1 - z_2 \vert^s \vert z_1 - z_3 \vert^t \vert z_1 - w \vert^u \vert z_2 - z_3 \vert^u \vert z_2 - w \vert^t
\vert z_3 - w \vert^s.
\ee
As $z_1, z_2$ and $z_3$ are on the real axis $\vert z_i - w \vert = \vert z_i - \bar{w} \vert$.

After fixing the twist operators to $(w, \bar{w}) = (i, -i)$ and one of the boundary vertex operators to $\infty$, we are typically
left with an integral of the form
\be
\label{GarousiMyersInt1}
I(a, b, c, d, e, f) = (2i)^f \int_{-\infty}^{\infty} dx_1 \int_{x_1}^{\infty} dx_2
(x_1 - i)^a (x_1 + i)^b (x_2 - i)^c (x_2 + i)^d (x_2 - x_1)^e.
\ee
As described in the appendix \ref{theusefulintegral}, (\ref{GarousiMyersInt1}) evaluates to
\bea
\label{GManalytic1}
I(a,b,c,d,e,f) & = & -(2i)^{3+a+b+c+d+e+f} \Gamma(-2-a-b-c-d-e) \ti \nn \\
& & \Big[ (-i)^{2(a+c)} \frac{\sin \pi(b+d+e) \Gamma(1+e) \Gamma(2+b+d+e) \Gamma(-1-d-e)}{\Gamma(-d)\Gamma(-a-c)} \ti \nn \\
& & \phantom{.}_3 F_2(-c,1+e,2+b+d+e;2+d+e,-a-c;1) \nn \\
& & + (-i)^{2(a+c+d+e)} \frac{\sin (\pi b) \Gamma(-1-c-d-e) \Gamma(1+b) \Gamma(1+d+e)}{\Gamma(-c) \Gamma(-1-a-c-d-e)} \ti \nn \\
& & \phantom{.}_3 F_2(-d,-1-c-d-e,1+b;-d-e,-1-a-c-d-e;1) \Big].
\eea

\subsection{Internal picture-changing}

Here we consider the picture-changing operator acting in the internal directions.
Contracting the boson $\mc{V}^{-1}(z_3)$ and one fermion vertex $\mc{V}^{-1/2}(z_2)$ with the picture-changing operator gives the following result:
\begin{align}
V_{\psi_2,int}^{(+\frac{1}{2})}(z_2) =& \ e^{+\frac{1}{2} \phi} \ S^{\mp} \ e^{i q^{\prime}_2 \cdot H} \ \pd \Zb_3 \ e^{i k_2 \cdot X}(z_2), \\
\nn V_{\phi,int}^{(0)}(z_3) =& \ \pd Z_3 \ e^{i k_3 \cdot X}(z_3),
\end{align}
where the internal H-charges are now $q_2^{\prime}=(-\frac{1}{2}, \frac{1}{2}, \frac{1}{2})$ and $q_3^{\prime}=(0,0,0)$.

The full amplitude can now be written as a product of correlators over superconformal ghosts, external spinors, internal spinors, momentum exponentials and bosonic twists:
\begin{align}
\mathcal{A}^{\textrm{int}}= & \langle e^{-\frac{1}{2} \phi (z_1)} e^{+\frac{1}{2} \phi(z_2) e^{- \phi (w)} e^{- \tilde{\phi}(\bar{w})}} \rangle \\
\nn & \langle e^{+ \frac{i}{2} (H_1+H_2)(z_1)} e^{- \frac{i}{2} (H_1+H_2)(z_2)} \rangle \\
\nn & \langle e^{i q_1 \cdot H(z_1)} e^{i q^{\prime}_2 \cdot H(z_2)} e^{i q_4 \cdot H(w)} e^{i q_5 \cdot H(\bar{w})}  \rangle \\
\nn & \langle e^{i k_1 \cdot X(z_1)} e^{i k_2 \cdot X(z_2)} e^{i k_3 \cdot X(z_3)} e^{i k_4 \cdot X(w)} e^{i k_4 \cdot X(\bar{w})}  \rangle \\
\nn & \langle \partial \Zb_3(z_2) \pd Z_3(z_3) \displaystyle\prod_{j=1}^3 \tw(w, \bar{w}) \rangle \ .
\end{align}
Using the basic CFT correlators from section 2 we can evaluate this amplitude.
The twist, ghost and fermionic correlators give
\bea
\label{lala}
& & (z_1 - z_2)^{-1} (z_2 - z_3)^{-2} (w - \bar{w})^{-2} (z_1 - w)^{-1+\theta_1} (z_1 - \bar{w})^{-\theta_1} \ti  \\
& &  (z_2 - w)^{\theta_2} (z_2 - \bar{w})^{-\theta_2} (z_3 - w)^{-1 + \theta_3} (z_3 - \bar{w})^{-\theta_3}
\left[ (z_3 - w)(z_2 - \bar{w}) - \theta_3 (z_3 - z_2)(w - \bar{w}) \right]. \nn
\eea
Including the external bosonic momentum correlators and the integral over the vertex operator locations, we have
\bea
\label{yodyo}
\int dz_1 \, dz_2 \, dz_3 \, d^2 w \, & & (z_1 - z_2)^{-1+s} (z_2 - z_3)^{-2 + u} (w - \bar{w})^{-2} (z_1 - w)^{-1+\theta_1+u/2} (z_1 - \bar{w})^{-\theta_1 +u/2} \ti \nn \\
& &  (z_2 - w)^{\theta_2 +t/2} (z_2 - \bar{w})^{-\theta_2 + t/2}
(z_3 - w)^{-1 + \theta_3 + s/2} (z_3 - \bar{w})^{-\theta_3 + s/2} \ti \nn \\
& & \left[ (z_3 - w)(z_2 - \bar{w}) - \theta_3 (z_3 - z_2)(w - \bar{w}) \right].
\eea
One can check that equation (\ref{yodyo}) is invariant under $SL(2, R)$ transformations.
Using the $SL(2, \mbb{R})$ symmetry, we first fix $(w, \bar{w}) \to (i, -i)$. This gives
\bea
\label{ofjoy}
\int dz_1 \, dz_2 \, dz_3  \, & & (z_1 - z_2)^{-1+s} (z_2 - z_3)^{-2 + u} (2i)^{-2} (z_1 - i)^{-1+\theta_1+u/2} (z_1 +i)^{-\theta_1 +u/2} \ti \nn \\
& &  (z_2 - i)^{\theta_2 +t/2} (z_2 + i)^{-\theta_2 + t/2}
(z_3 - i)^{-1 + \theta_3 + s/2} (z_3 + i)^{-\theta_3 + s/2} \ti \nn \\
& & \left[ (z_3 - i)(z_2 + i) - 2 i \theta_3 (z_3 - z_2) \right].
\eea

To consider all contributions to the amplitude we need to include all orderings of vertex operators on the boundary with respect to the branch point as set out in section (\ref{branchpt}). Although there are six orderings of three vertex operators and the branch point that are inequivalent on the disk we only need to consider the three configurations corresponding to the same cyclic ordering of open string vertex operators on the boundary (the anticyclic ordering vanishes due to the Chan-Paton traces).
Hence we only consider the three separate ways of fixing vertex operators for the cyclic ordering $z_1 z_2 z_3$: we either set $z_3 \to \infty$, $z_2 \to \infty$ or $z_1 \to \infty$. In the absence of a twist field, these would be identical, but
as described in section \ref{branchpt} they will produce distinct results.

Fixing $z_3 \to \infty$ and including the c-ghost correlation function $\langle c(z_3) c(w) c(\bar{w}) \rangle = (2i)(\infty)^2$, we obtain
\bea
\frac{-1}{2i} \int_{-\infty}^{+\infty} dz_1 \int_{z_1}^{+\infty} dz_2 & & (z_2 - z_1)^{-1+s} (z_1 - i)^{-1+\theta_1 +u/2}
(z_1 + i)^{-\theta_1 + u/2} (z_2 - i)^{\theta_2 + t/2} (z_2 + i)^{-\theta_2 + t/2} \nn \\
& & \ti \left[ (z_2 + i) - 2i \theta_3 \right].
\eea
We can evaluate this using the standard integral $I(a, b, c, d, e, f)$ given by \eqref{GManalytic1}, and in a similar vein we obtain the results
for $z_2 \to \infty$ and $z_1 \to \infty$. We find:
\bea
\label{inin1}
z_3 \to \infty & : &  -I \Big(-1 + \theta_1 + u/2, - \theta_1 + u/2, \theta_2 + t/2, 1 - \theta_2 + t/2, -1 + s, -1 \Big) \\
& & + 2i \theta_3 \ti I \Big( -1+\theta_1 + u/2, - \theta_1 + u/2, \theta_2 +t/2, -\theta_2+t/2, -1+s, -1 \Big). \nn \\
\label{inin2}
z_2 \to \infty & : & -I \Big(\theta_3 + s/2, - \theta_3 + s/2, -1+\theta_1 +u/2, -\theta_1+u/2, t, -1 \Big)   \\
& & - 2 i \theta_3 I\Big( -1+\theta_3 + s/2, -\theta_3 + s/2, -1+\theta_1 + u/2, -\theta_1 + u/2, t, -1 \Big). \nn \\
\label{inin3}
z_1 \to \infty & : & I \Big(\theta_2 + t/2,1 - \theta_2 + t/2, \theta_3 +s/2, -\theta_3+s/2, -2+u, -1 \Big)   \\
& & -2i\theta_3 I\Big( \theta_2 + t/2, -\theta_2 + t/2, -1+\theta_3 + s/2, -\theta_3 + s/2, -1+u, -1 \Big). \nn
\eea

\subsection{External picture-changing}

The external picture changing leads to two distinct contributions which must be computed separately. They differ by the structure of
H-charges that arise.

\subsubsection*{Case 1}

In this case the H-charges take the form
\begin{align}
V_{\psi_1}^{(-\frac{1}{2})}(z_1) \ \sim & \ e^{-\phi/2}|+ \, +\rangle \otimes  |+ \, - \, -\rangle \\
\nn V_{\psi_2}^{(-\frac{1}{2})}(z_2) \ \sim & \ e^{\phi/2}|(---) \,-\rangle \otimes  |- \, +\, -\rangle \\
\nn V_{\phi}^{(-1)}(z_3) \ \sim & \ |(++), \, 0\rangle \otimes  |0 \, 0 \, (++) \rangle \\
\nn V_{\textrm{tw}}^{(-1)}(w) \ \sim & \ e^{-\phi}|00\rangle \otimes  |\theta_1, \ \theta_2, \ \theta_3 \rangle \\
\nn V_{\textrm{tw}}^{(-1)}(\bar{w}) \ \sim & \ e^{-\tilde{\phi}}|00\rangle \otimes  |- \theta_1, \ -\theta_2, \ -\theta_3 \rangle \ .
\end{align}
As $e^{\phi(z)} e^{-\phi(w)/2} \sim (z-w)^{\half} e^{\phi/2}$, $\, e^{-iH(z)}e^{-iH(w)/2} \sim (z-w)^{\half}e^{-3iH(w)/2}$
and $\partial X^I(z) e^{ik_2 \cdot X(w)} \sim \frac{ik_2^I}{(z-w)} e^{ik_2 \cdot X(w)}$, the picture changing of $V_{\psi_2}(z_2)$
is unambiguous and there is no need to consider subleading terms in the OPE. The momentum factors in this amplitude
are given by $k_3^{1-} k_2^{1+}$. As $k_2^{1+}$ is the only non-zero component of $k_2$, we can promote this to $k_2 \cdot k_3 = u$.

Evaluating the twist, ghost and fermionic operators, we obtain
\bea
& & -(w - \bar{w})^{-2} (z_1 - z_2)^{-1} (z_2 - z_3)^{-2} (z_1 - w)^{-1 + \theta_1} (z_1 - \bar{w})^{-\theta_1} \\
& & \ti (z_2 - w)^{\theta_2} (z_2 - \bar{w})^{1-\theta_2} (z_3 - w)^{\theta_3} (z_3 - \bar{w})^{-\theta_3} \nn
\eea
There is an overall minus sign in this expression (compared to the analogous expressions (\ref{lala}) for internal picture changing and
(\ref{rocky}) for external picture changing case 2). This arises as the picture changing operator introduces negative H-charge for $V(z_2)$ and positive H-charge for $V(z_3)$.
This is equivalent to introducing a correlator $\bar{\psi}(z_2) \psi(z_3)$, whereas the other two cases
introduce a correlator $\psi(z_2) \bar{\psi}(z_3)$, leading to the minus sign differential.

Prior to fixing the $SL(2, \mbb{R})$ symmetry, the amplitude is given by
\bea
\label{yodyo1}
& -u \int \, dz_1 \, dz_2 \, dz_3 \, d^2 w \, & (w - \bar{w})^{-2} (z_1 - z_2)^{-1+s} (z_2 - z_3)^{-2+u} (z_1 - w)^{-1 + \theta_1+u/2} (z_1 - \bar{w})^{-\theta_1+u/2} \nn \\
& & \ti (z_2 - w)^{\theta_2+t/2} (z_2 - \bar{w})^{1-\theta_2+t/2} (z_3 - w)^{\theta_3+s/2} (z_3 - \bar{w})^{-\theta_3+s/2}
\eea
It is easy to check invariance of (\ref{yodyo1}) under $SL(2, \mbb{R})$ transformations. We then fix $(w, \bar{w}) \to (i, -i)$, giving
\bea
& \frac{-u}{(2i)^2} \int \, dz_1 \, dz_2 \, dz_3 \, &  (z_1 - z_2)^{-1+s} (z_2 - z_3)^{-2+u} (z_1 - i)^{-1 + \theta_1+u/2}
(z_1 + i)^{-\theta_1+u/2} \nn \\
& & \ti (z_2 - i)^{\theta_2+t/2} (z_2 + i)^{1-\theta_2+t/2} (z_3 - i)^{\theta_3+s/2} (z_3 + i)^{-\theta_3+s/2}
\eea
This is precisely the same integral as arose in eq. (\ref{ofjoy}) in
the analysis of internal picture changing. We can then immediately write down the result
for the three separate cases of $z_3 \to \infty$, $z_2 \to \infty$ and $z_1 \to \infty$. These are
\bea
\label{xx1}
z_3 \to \infty & : & u \ti I(-1+\theta_1 + u/2, -\theta_1 + u/2, \theta-2+t/2, 1-\theta_2 + t/2, -1+s,-1), \\
\label{xx2}
z_2 \to \infty & : & u \ti I(\theta_3 + s/2, -\theta_3 + s/2, -1+\theta_1 + u/2, -\theta_1 + u/2, t, -1), \\
\label{xx3}
z_1 \to \infty & : & -u \ti I(\theta_2 + t/2, 1-\theta_2 + t/2, \theta_3 + s/2, -\theta_3 + s/2, -2 + u, -1).
\eea

\subsubsection*{Case 2}

In this case the H-charges take the form
\begin{align}
V_{\psi_1}^{(-\frac{1}{2})}(z_1) \ \sim & \ e^{-\phi/2}|+ \, +\rangle \otimes  |+ \, - \, -\rangle \\
\nn V_{\psi_2}^{(-\frac{1}{2})}(z_2) \ \sim & \ e^{\phi/2}|+ \,-\rangle \otimes  |- \, +\, -\rangle \\
\nn V_{\phi}^{(-1)}(z_3) \ \sim & \ |(--), \, 0\rangle \otimes  |0 \, 0 \, (++) \rangle \\
\nn V_{\textrm{tw}}^{(-1)}(w) \ \sim & \ e^{-\phi}|0 \, 0\rangle \otimes  |\theta_1, \ \theta_2, \ \theta_3 \rangle \\
\nn V_{\textrm{tw}}^{(-1)}(\bar{w}) \ \sim & \ e^{-\tilde{\phi}}|0 \, 0\rangle \otimes  |- \theta_1, \ -\theta_2, \ -\theta_3 \rangle \ .
\end{align}
The momentum prefactor is $k_3^{1+}$.
Here the picture-changing is more subtle as
\bea
e^{\phi(z)} e^{-\phi(w)/2} & \sim & (z-w)^{\half} e^{\phi(w)/2}, \nn \\
e^{iH(z)}e^{-iH(w)/2} & \sim & (z-w)^{-\half} e^{iH(w)/2}, \nn \\
\partial X^I(z) e^{ik_2 \cdot X(w)} & \sim & \frac{ik_2^I}{(z-w)} e^{ik_2 \cdot X(w)} + \partial X^I(w) e^{ik_2 \cdot X(w)}.
\eea
The leading term in the OPE is at $\mc{O}(z-w)^{-1}$, whereas we require the term at $\mc{O}(z-w)^{0}$. In principle we should expand
the ghost, fermionic and bosonic OPEs to obtain the $\mc{O}(z-w)^{0}$ term. However in fact only the subleading bosonic term is relevant.
The subleading ghost and fermionic terms necessarily involve the leading bosonic term, which has a factor of
$k_2^{1-}$ and so vanishes identically.

The ghost and fermion correlators give
\bea
\label{rocky}
& & (w-\bar{w})^{-2}(z_1 - z_3)^{-1} (z_2 - z_3)^{-1} (z_1 - w)^{-1 + \theta_1} (z_1 - \bar{w})^{-\theta_1} \\
& & \ti (z_2 - w)^{\theta_2} (z_2 - \bar{w})^{1-\theta_2} (z_3 - w)^{\theta_3} (z_3 - \bar{w})^{-\theta_3}. \nn
\eea
The bosonic correlator is
\be
k_3^{1+} \langle e^{ik_1 \cdot X(z_1)} \partial_t X^{1-} (z_2) e^{i k_2 \cdot X(z_2)} e^{i k_3 \cdot X(z_3)} e^{i k_4 \cdot X(w, \bar{w})} \rangle.
\ee
This gives
\bea
k_3^{1+} \left( \frac{k_1^{1-}}{(z_2 - z_1)} + \frac{k_4^{1-}}{2 (z_2 - w)} + \frac{k_4^{1-}}{2(z_2 - \bar{w})} \right)
& \vert z_1 - z_2 \vert^s \vert z_1 - z_3 \vert^t \vert z_1 - w \vert^u & \nn \\
& \ti \vert z_2 - z_3 \vert^u \vert z_2 - w \vert^t
\vert z_3 - w \vert^s.  &
\eea
We have dropped the $k_3^{1+} k_3^{1-}$ term as it Lorentz completes into $k_3^2 = 0$. Using $k_4^{1-} = - k_1^{1-}
- k_2^{1-} -k_3^{1-}$, this effectively becomes
$$
k_3^{1+} \left( \frac{k_1^{1-}}{(z_2 - z_1)} - \frac{k_1^{1-}}{2 (z_2 - w)} - \frac{k_1^{1-}}{2(z_2 - \bar{w})} \right)
\vert z_1 - z_2 \vert^s \vert z_1 - z_3 \vert^t \vert z_1 - w \vert^u \vert z_2 - z_3 \vert^u \vert z_2 - w \vert^t
\vert z_3 - w \vert^s.
$$
As $k_1^{1-}$ is the only non-zero component of $k_1$, we can promote $k_3^{1+} k_1^{1-}$ to $k_3 \cdot k_1 = t$.
The amplitude then becomes
\bea
\label{yodyo2}
t \, \int \, dz_1 \, dz_2 \, dz_3 \, d^2 w \, & & (w - \bar{w})^{-2} (z_1 - z_3)^{-1+t} (z_2 - z_3)^{-1+u} (z_1 - w)^{-1 + \theta_1+u/2} (z_1 - \bar{w})^{-\theta_1+u/2} \nn \\
& & \ti (z_2 - w)^{\theta_2+t/2} (z_2 - \bar{w})^{1-\theta_2+t/2} (z_3 - w)^{\theta_3+s/2} (z_3 - \bar{w})^{-\theta_3+s/2} \nn \\
& & \ti \left( \frac{1}{(z_2 - z_1)} - \frac{1}{2} \left( \frac{1}{(z_2 - w)} + \frac{1}{(z_2 - \bar{w})} \right) \right).
\eea
One can again check that this is $SL(2, \mbb{R})$ invariant.
We fix $(w, \bar{w}) \to (i, -i)$ and consider the three separate cases $z_3 \to \infty$, $z_2 \to \infty$ and $z_1 \to \infty$. The three
results are then
\bea
\label{yy1}
z_3 \to \infty & : & t \ti I \Big(-1 + \theta_1 + u/2, - \theta_1 + u/2, \theta_2 + t/2, 1 - \theta_2 + t/2, -1 + s, -1 \Big) \\
& & -\frac{t}{2} \ti I \Big( -1+\theta_1 + u/2, - \theta_1 + u/2, -1+\theta_2 +t/2, 1-\theta_2+t/2, s, -1 \Big)  \nn \\
& & -\frac{t}{2} \ti I\Big( -1+\theta_1 + u/2, -\theta_1 + u/2, \theta_2 + t/2, -\theta_2 + t/2, s, -1 \Big). \nn \\
\label{yy2}
z_2 \to \infty & : & \frac{t}{2} \ti I \Big(\theta_3 + s/2, - \theta_3 + s/2, \theta_1 +u/2, -\theta_1+u/2, -1+t, -1 \Big)   \\
& & +\frac{t}{2} \ti I\Big( \theta_3 + s/2, -\theta_3 + s/2, -1+\theta_1 + u/2, 1-\theta_1 + u/2, -1+t, -1 \Big). \nn \\
\label{yy3}
z_1 \to \infty & : & \frac{t}{2} \ti I \Big(-1+\theta_2 + t/2,1 - \theta_2 + t/2, \theta_3 +s/2, -\theta_3+s/2, -1+u, -1 \Big)  \\
& & +\frac{t}{2} \ti I\Big( \theta_2 + t/2, -\theta_2 + t/2, \theta_3 + s/2, -\theta_3 + s/2, -1+u, -1 \Big) \nn .
\eea

\subsection{Combination}

To obtain a complete amplitude we must combine the three picture-changing contributions considered above. These then give
complete expressions for the three separate ways of fixing vertex operators, $z_3 \to \infty$, $z_2 \to \infty$ and $z_1 \to \infty$:
\bea
\label{Bigexp1}
\mc{A}_{z_3 \to \infty} &: & (u+t) \ti I \Big(-1 + \theta_1 + u/2, - \theta_1 + u/2, \theta_2 + t/2, 1 - \theta_2 + t/2, -1 + s, -1 \Big) \nn \\
& & -\frac{t}{2} \ti I \Big( -1+\theta_1 + u/2, - \theta_1 + u/2, -1+\theta_2 +t/2, 1-\theta_2+t/2, s, -1 \Big)  \nn \\
& & -\frac{t}{2} \ti I\Big( -1+\theta_1 + u/2, -\theta_1 + u/2, \theta_2 + t/2, -\theta_2 + t/2, s, -1 \Big) \\
& & -I \Big(-1 + \theta_1 + u/2, - \theta_1 + u/2, \theta_2 + t/2, 1 - \theta_2 + t/2, -1 + s, -1 \Big) \nn \\
& & + 2i \theta_3 \ti I \Big( -1+\theta_1 + u/2, - \theta_1 + u/2, \theta_2 +t/2, -\theta_2+t/2, -1+s, -1 \Big). \nn \\
& & \nn \\
\label{Bigexp2}
\mc{A}_{z_2 \to \infty} & : & \frac{t}{2} \ti I \Big(\theta_3 + s/2, - \theta_3 + s/2, \theta_1 +u/2, -\theta_1+u/2, -1+t, -1 \Big) \nn   \\
& & +\frac{t}{2} \ti I\Big( \theta_3 + s/2, -\theta_3 + s/2, -1+\theta_1 + u/2, 1-\theta_1 + u/2, -1+t, -1 \Big) \nn \\
& & + u \ti I \Big(\theta_3 + s/2, -\theta_3 + s/2, -1+\theta_1 + u/2, -\theta_1 + u/2, t, -1 \Big) \\
& & -I \Big(\theta_3 + s/2, - \theta_3 + s/2, -1+\theta_1 +u/2, -\theta_1+u/2, t, -1 \Big)  \nn \\
& & - 2 i \theta_3 I\Big( -1+\theta_3 + s/2, -\theta_3 + s/2, -1+\theta_1 + u/2, -\theta_1 + u/2, t, -1 \Big). \nn \\
& & \nn \\
\label{Bigexp3}
\mc{A}_{z_1 \to \infty} & : &  \frac{t}{2} \ti I \Big(-1+\theta_2 + t/2,1 - \theta_2 + t/2, \theta_3 +s/2, -\theta_3+s/2, -1+u, -1 \Big)  \nn \\
& & +\frac{t}{2} \ti I\Big( \theta_2 + t/2, -\theta_2 + t/2, \theta_3 + s/2, -\theta_3 + s/2, -1+u, -1 \Big) \nn \\
& & -u \ti I(\theta_2 + t/2, 1-\theta_2 + t/2, \theta_3 + s/2, -\theta_3 + s/2, -2 + u, -1) \\
& & + I \Big(\theta_2 + t/2,1 - \theta_2 + t/2, \theta_3 +s/2, -\theta_3+s/2, -2+u, -1 \Big) \nn  \\
& & -2i\theta_3 I\Big( \theta_2 + t/2, -\theta_2 + t/2, -1+\theta_3 + s/2, -\theta_3 + s/2, -1+u, -1 \Big). \nn
\eea
We want to consider each of these in the limit $s, t, u \to 0$ with $s+t+u=0$.
There is a powerful consistency check on these expressions. Each of the individual terms (\ref{inin1}) to (\ref{inin3}),
(\ref{xx1}) to (\ref{xx3}), and (\ref{yy1}) to (\ref{yy3}) that appear above have an unphysical pole of the form $(s+t+u)^{-1}$.
These are incompatible with the known structure of either field theory or string theory. However on combination
all such unphysical poles vanish: each of (\ref{Bigexp1}) to (\ref{Bigexp3}) is well-behaved in the limit $s+t+u \to 0$ and
only has poles in $\frac{1}{s}$, $\frac{1}{t}$ or $\frac{1}{u}$.
These checks are easiest to carry out numerically as
it is cumbersome to treat the hypergeometric functions
analytically.

$\mc{A}_{z_3 \to \infty}$, $\mc{A}_{z_2 \to \infty}$ and $\mc{A}_{z_1 \to \infty}$ are all distinct functions of $s$, $t$ and $u$.
At first this may seem surprising as they each arise from the same $SL(2, \mbb{R})$ invariant expression. However there is no contradiction,
and this originates from the presence of the twist operator as explained in section \ref{branchpt}.

\begin{figure}
\begin{center}
\includegraphics[height=5cm]{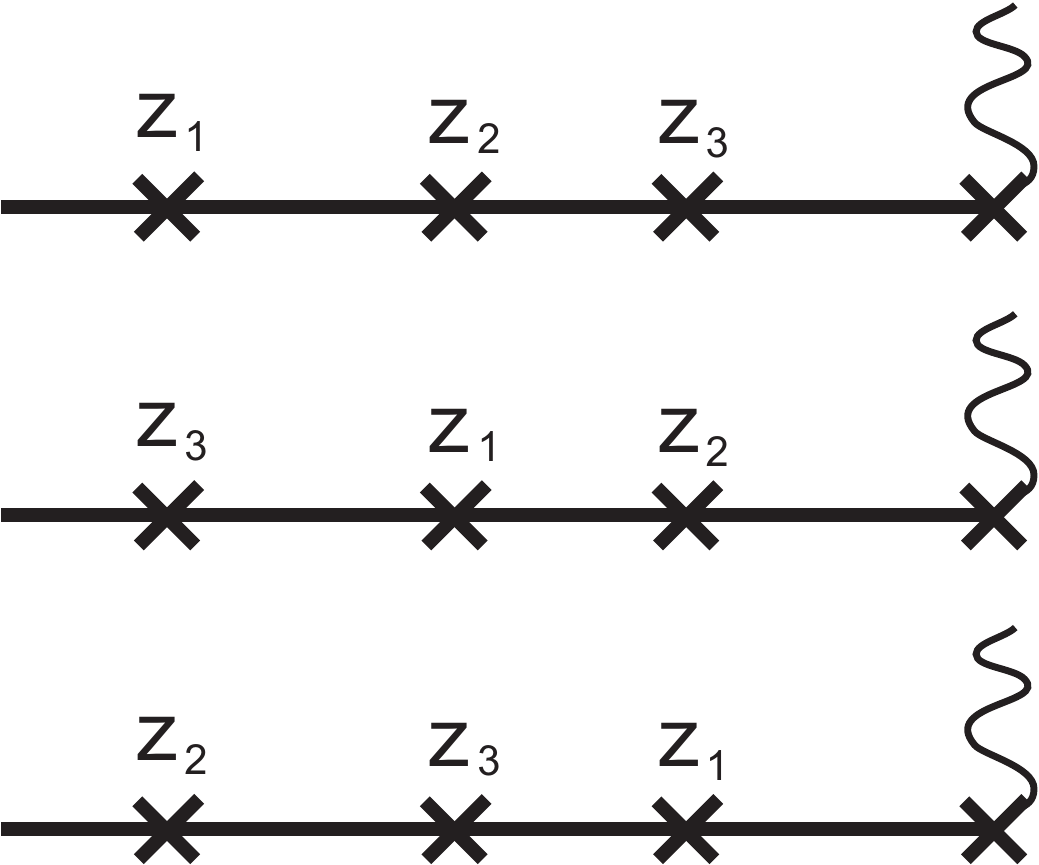}
\caption{The three possible (and distinct) orderings of the vertex operator in relation to the location of the branch point
on the boundary of the disk.}
\label{VOOrderings1}\end{center}\end{figure}

We review how the insertion of a twist operator affects the disk calculation based on our analysis of a Yukawa coupling. The twist operator introduces a branch cut into the correlation functions of the boundary vertex operators. The necessary existence
of this branch cut is manifest in figure \ref{VOOrderings1}: as the vertex operators must shift by a phase on translation around the twist operator - equivalently
on translation around the boundary - the correlation function must have a branch point. In equations (\ref{yodyo}), (\ref{yodyo1}) and
 (\ref{yodyo2}) we have always chosen this branch
point to be at infinity (implicitly, by using the conventional definition of $z^{\theta}$ for $0 < \theta < 1$) The presence of the branch point implies there are three distinct configurations per cyclic ordering of the vertex operators and we need to sum over all configurations. We only need to examine the three orderings shown in figure \ref{VOOrderings1} as the anticyclic ordering $(z_1, z_3, z_2)$ vanishes due to the Chan-Paton factors.

A little care is needed in using the $SL(2, \mbb{R})$ symmetry to fix the location of vertex operators. Having fixed $(w, \bar{w}) \to
(i, -i)$ the residual transformation is of the form
$$
z \to \frac{a z + b}{-b z + a}, \qquad a, b \in \mbb{R}, \quad a^2 + b^2 = 1.
$$
However in general such a transformation also moves the branch point location away from $\infty$, whereas in our fixed expressions
we wish to keep the location of the branch point at infinity.

\begin{figure}
\begin{center}
\includegraphics[height=5cm]{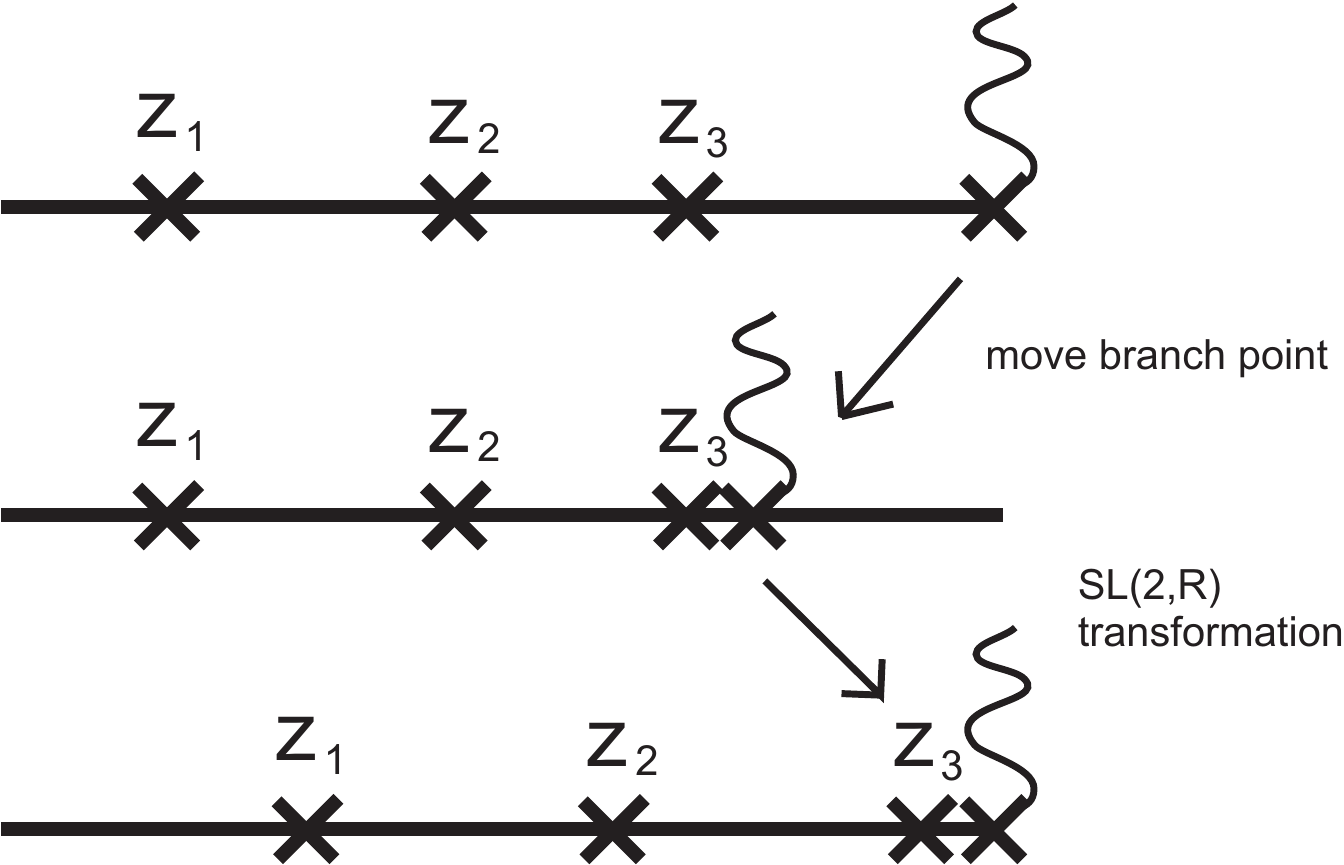}
\caption{The procedure of fixing the $SL(2, \mbb{R})$ degeneracy while keeping the branch point at $\infty$.}
\label{VOOrderings}\end{center}\end{figure}

Amplitudes are continuous except when a vertex operator moves through a branch point location. Furthermore, the overall amplitude
(summing over all possible orderings) must be insensitive to the location of the branch point.
For each operator ordering, the $SL(2, \mbb{R})$ degeneracy is then dealt with as follows.
\begin{enumerate}
\item
For any set of values $(z_1, z_2, z_3)$, the amplitude does not change unless the branch point is moved through a vertex operator.
We can use this freedom to bring the branch point from $\infty$ so that it resides next to $z_3$.
\item
We can now use an $SL(2, \mbb{R})$ transformation to take $z_3 \to \infty$. As the branch point is next to $z_3$, the branch point is also
moved to $\infty$, and so our previous $SL(2,\mbb{R})$-fixed expressions with the branch point at $\infty$ remain valid.
\item
Repeat for the orderings $(z_3, z_1, z_2)$ and $(z_2, z_3, z_1)$.
\end{enumerate}
This is illustrated in figure \ref{VOOrderings}.

We finally need to combine all different orderings. To do so amplitudes have to be well-defined. By themselves the monodromy from the
twist operator would prevent this (the amplitude with a vertex operator moved around the disk would change by a factor $e^{2 \pi i \theta}$).
To make the amplitudes well-defined, we need to introduce a compensating phase of $e^{-2 \pi i \theta}$ whenever a vertex operator moves through the
branch point. This phase can alternatively be derived by modifying the Chan-Paton trace as
\be
\hbox{Tr}(T_1 T_2 T_3) \to \hbox{Tr}(T_1 T_2 T_3 \gamma_{\theta}).
\ee
Here $\gamma_{\theta}$ is understood to be inserted at the boundary at the location of the branch point.
As the Chan-Paton matrices do not commute with the twist matrix $\gamma_{\theta}$, $T_i \gamma_{\theta} = e^{2 \pi i \theta_i} \gamma_{\theta} T_i$
we similarly obtain an extra phase when vertex operators are moved through the branch point.
There is also an extra factor of $(-1)$ that occurs as either $\mc{V}(z_1)$ or $\mc{V}(z_2)$ is moved through the branch point. This factor
is due to our use of conventions with $\theta_1 + \theta_2 + \theta_3 = 1$, and comes from the
phase $e^{\pi i(\theta_1 + \theta_2 + \theta_3)}$.\footnote{This is easiest to see by considering the case of a gaugino vertex operator, with H-charges labelled by $(+,+,+,+,+)$ on a boundary
with the twist field in the interior. The gaugino belongs to the  untwisted sector and so should have no monodromy about the twist field. However it is easy to see from the H-charges that it does have a monodromy of $e^{i \pi (\theta_1 + \theta_2 + \theta_3)}$ on movement around the disk. In conventions where $\theta_1 + \theta_2 + \theta_3 = 1$ (rather than 0) we then need an additional minus sign appearing as spacetime spinors are moved through the branch cut.}

The appropriate amplitudes to combine are then
$\mc{A}_{z_3 \to \infty}$, $-e^{-2 \pi i \theta_1} \mc{A}_{z_1 \to \infty}$ and $e^{-2 \pi i (\theta_1 + \theta_2)} \mc{A}_{z_2 \to \infty}$.
Summing over all orderings, the full amplitude is then
\be
\label{first}
\mc{A}^{'}_{\textrm{full}} = \mc{A}_{z_3 \to \infty} - e^{-2 \pi i \theta_1} \mc{A}_{z_1 \to \infty} + e^{-2 \pi i (\theta_1 + \theta_2)} \mc{A}_{z_2 \to \infty}.
\ee
The amplitudes $\mc{A}_{z_3 \to \infty}$, $-e^{-2 \pi i \theta_1} \mc{A}_{z_1 \to \infty}$ and $e^{-2 \pi i (\theta_1 + \theta_2)} \mc{A}_{z_2 \to \infty}$ have a remarkable pole structure, summarised below:
\bea
\mc{A}_{z_3 \to \infty} & = & \frac{\alpha_s}{s} + \frac{\alpha_t}{t} + \frac{\alpha_u}{u}, \nn \\
e^{-2 \pi i (\theta_1 + \theta_2)}\mc{A}_{z_2 \to \infty} & = & \frac{\beta_s}{s} \phantom{ + \frac{\beta_t}{t} + } + \frac{\beta_u}{u}, \nn \\
-e^{-2 \pi i \theta_1}\mc{A}_{z_1 \to \infty} & = & \frac{\gamma_s}{s} \, + \frac{\gamma_t}{t} \, \, + \frac{\gamma_u}{u},
\eea
where $\beta_s = -\gamma_s$, $\alpha_t + \gamma_t = 0$ and $\alpha_u - \beta_u + \gamma_u = 0$. The explicit expressions for the coefficients
 $\alpha$, $\beta$ and $\gamma$ in terms of the angles $(\theta_1, \theta_2, \theta_3)$ are given in appendix \ref{poles}.

\begin{figure}
\begin{center}
\includegraphics[height=4cm]{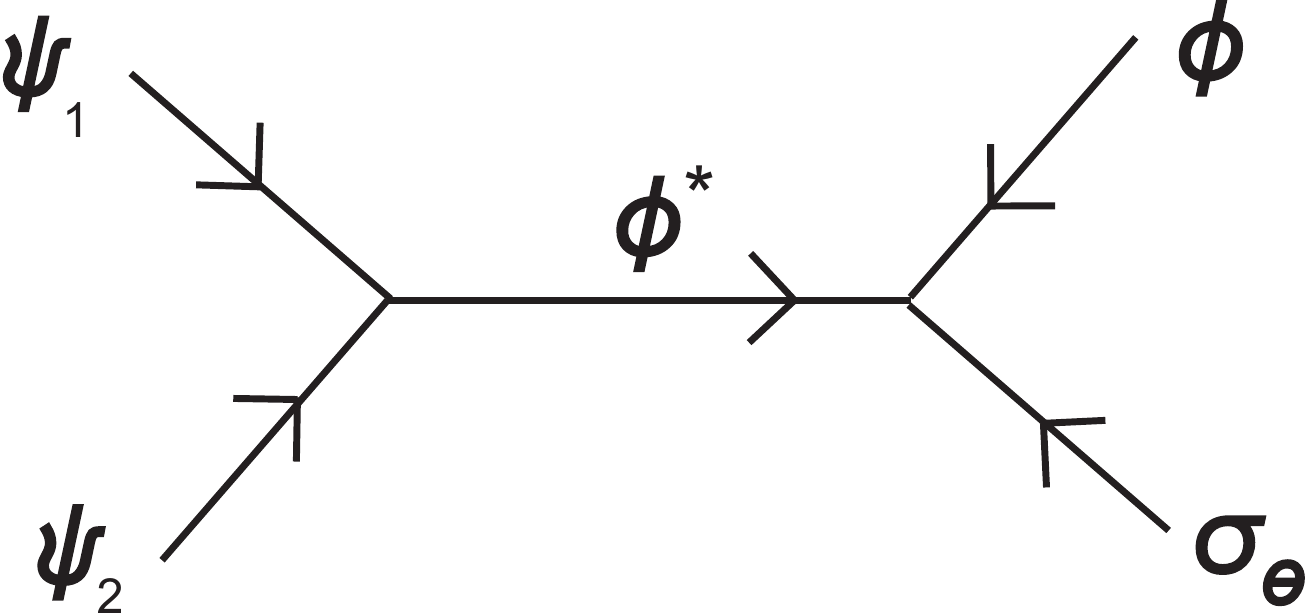}
\caption{The origin of the $1/s$ pole as factorisation of the 4-point diagram onto a 3-pt Yukawa and an FI term.}
\label{YukawaFact}\end{center}\end{figure}

We see that the $\frac{1}{t}$ pole cancels and, up to an overall sign in $\mc{A}_{z_2 \to \infty}$, so does the $\frac{1}{u}$ pole.
The low-energy supergravity theory appears to require both the presence of a $\frac{1}{s}$ pole
and the absence of poles in $\frac{1}{u}$ or $\frac{1}{t}$. The $\frac{1}{s}$ pole
comes from the field theory diagram shown in figure \ref{YukawaFact}, factorising onto the 3-point Yukawa interaction and the
Fayet-Iliopolous D-term $(\phi \phi^{*} + \xi)^2$, with $\xi \sim \tau_{blowup}$.

The cancellation present strongly suggest that an overall sign is missing
in (\ref{first}) in the computation of $\mc{A}_{z_2 \to \infty}$. This sign (which can be written as $\hbox{sign}(z_3 - z_2)$)
is presumably due to a cocycle factor that is present in the ordering of the vertex operators.
We shall work on the supposition that the correct expression is
indeed
\be
\label{Afull}
\mc{A}_{\textrm{full}} = \mc{A}_{z_3 \to \infty} - e^{-2 \pi i \theta_1} \mc{A}_{z_1 \to \infty} - e^{-2 \pi i (\theta_1 + \theta_2)} \mc{A}_{z_2 \to \infty}.
\ee
In this case the surviving momentum pole of the full result is:
\be
\label{Afullpole}
\mc{A}_{\textrm{full}} = \frac{i e^{\pi i \theta_3} \sin{\pi \theta_3}}{s} + \ldots.
\ee
and corresponds to the diagram in figure \ref{YukawaFact}. We can exploit this factorisation channel for a consistency-check of our expression: the diagram consists of a tree-level Yukawa coupling which is proportional to a constant; the other vertex arises from the tree-level coupling between charged matter and a single blow-up mode $\langle C_3 \bar{C}_3 b\rangle$ which we calculated in section \ref{mattermetric}. There the leading term of our result \eqref{CCb} for small momenta is $i e^{\pi i \theta_3} \sin(\pi \theta_3)$ and it is exactly this combination that also appears in our result for the factorisation limit of the 4-point function. We interpret this agreement as firm support for our result \eqref{Afull}.

We now present strong further evidence for this. Instead of working with $\theta_1 + \theta_2 + \theta_3 = 1$, we can equally well work with
$\theta_1 + \theta_2 + \theta_3 = 0$. This can be accomplished by taking $\theta_1 \to \theta_1 - 1$ or $\theta_2 \to \theta_2 - 1$.
The only place the angles $\theta_1$ or $\theta_2$ appear is in the fermionic correlators. These are divorced from any subtleties
involving the bosonic twist fields and it is easy to work out the appropriate modification to the correlation functions. The result is that
for $\theta_1 \to \theta_1 - 1$, we should multiply the amplitudes by
$$
(z_1 - w)^{-1/2}(z_1 - \bar{w})^{\half} (z_2 - w)^{\half} (z_2 - \bar{w})^{-\half}.
$$
For $\theta_2 \to \theta_2 - 1$, we should multiply the amplitudes by
$$
(z_1 - w)^{1/2}(z_1 - \bar{w})^{-\half} (z_2 - w)^{-\half} (z_2 - \bar{w})^{\half}.
$$
The modifications to equations (\ref{Bigexp1}), (\ref{Bigexp2}) and (\ref{Bigexp3}) are easy to determine. Let us denote each of the integrals
appearing in these equations by $I(a, b, c, d, e, f)$. Then for $\theta_1 \to \theta_1 - 1$ the appropriate modifications are
\bea
z_3 \to \infty & : & a \to a - \half, \quad b \to b + \half, \quad c \to c + \half, \quad  d \to d - \half, \nn \\
z_2 \to \infty & : & c \to c - \half, \quad  d \to d + \half,  \\
z_1 \to \infty & : & a \to a + \half, \quad  b \to b - \half. \nn
\eea
For $\theta_2 \to \theta_2 - 1$ the appropriate modifications are
\bea
z_3 \to \infty & : & a \to a + \half, \quad b \to b - \half, \quad c \to c - \half, \quad d \to d + \half, \nn \\
z_2 \to \infty & : & c \to c + \half, \quad d \to d - \half,  \\
z_1 \to \infty & : & a \to a - \half, \quad b \to b + \half. \nn
\eea
These amplitudes should be combined as
\be
\label{Afull2}
\mc{A}_{\textrm{full}}^{\Delta \theta = 1} = \mc{A}^{\Delta \theta = 1}_{z_3 \to \infty}
- e^{-2 \pi i (\theta_1 + \theta_2)} \mc{A}^{\Delta \theta = 1}_{z_2 \to \infty}
+ e^{-2 \pi i \theta_1} \mc{A}^{\Delta \theta = 1}_{z_1 \to \infty}.
\ee
Again we find that the $\frac{1}{t}$ pole cancels automatically and the $\frac{1}{u}$ pole cancels once we have introduced the
relative minus sign in $\mc{A}^{\Delta \theta = 1}_{z_2 \to \infty}$ in (\ref{Afull2}).

A strong check on these expressions is that, although e.g. $\mc{A}^{\Delta \theta = 1}_{z_3 \to \infty}$
and $\mc{A}_{z_3 \to \infty}$ have very different individual values, the combined sums (\ref{Afull}) and
(\ref{Afull2}) are identical. This is as required if these are to represent physical amplitudes. Furthermore,
this holds if and only if the extra relative minus sign in $\mc{A}_{z_2 \to \infty}$ is introduced. We therefore conclude that
the minus sign introduced in equations (\ref{Afull}) and (\ref{Afull2}) is necessary to obtain correct physical amplitudes,
and leave the origin of the sign for future work.

Our motivation for performing this calculation was to determine whether contact terms of the form $\int {\textrm{d}}^4 x \tau_s \psi \psi \phi$ exist in the effective action. To answer this question we have to go beyond the momentum pole: a non-zero contribution to our result for vanishing momenta beyond the pole is evidence for the presence of such a contact term. We do have an implicit analytic expression for this term; it is given by our result for the full amplitude with the pole subtracted in the limit that $s$, $t$ and $u$ vanish:
\be
\mathcal{A}_{\textrm{finite}} = \lim_{\substack{s \rightarrow 0 \\ t \rightarrow 0 \\ u \rightarrow 0 \\}} \left(\mathcal{A}_{\textrm{full}} - \frac{i e^{\pi i \theta_3} \sin{\pi \theta_3}}{s} \right)
\ee
Although it would be very cumbersome to obtain the result in this limit analytically, one can nevertheless show numerically that a non-zero limit exists which also depends on the orbifold twist angles $\theta_i$. We can conclude the following from this result: in the case that the twist field is located at the same orbifold singularity as the stack of D3-branes, the quantum correlator gives the full result for the amplitude of interest. The existence of a finite result in the zero-momentum limit then confirms the existence of the contact interaction $\int {\textrm{d}}^4 x \tau_s \psi \psi \phi$. Yet, if the twist is geometrically separated from the D3-branes the amplitude will include a non-trivial contribution from the classical embedding of the spacetime coordinates and which will have an important impact on the result. We study these classical contributions in the next section and extend the discussion to the insertion of multiple twist fields.

\section{Distant Twist Fields}

We now return to consider correlation functions involving twist fields that are physically separated from the D3-brane stack.
In this case in addition to the quantum correlators considered above we also need a classical correlator. In this section we revert to denoting the complexified bosonic coordinates by $X_i$ instead of $Z_i$ to avoid confusion with worldsheet coordinates.

Recall that a D3-brane located at $X_{i} = x_{i}$ imposes the disk boundary condition
\be
X_i(\hbox{Im}(z) = 0) = x_i.
\ee
In space-time any given twist field is associated to an orbifold singularity located at
$(x^{tw}_1, x^{tw}_2, x^{tw}_3)$. Inserting this closed string twist field at the world sheet location $z_0$, $\zb_0$
implies that the worldsheet embedding $X(z, \zb)$ should satisfy the following conditions:
\bea
\label{yoda2}
X_i(z_0, \zb_0) & = & (x^{tw}_1, x^{tw}_2, x^{tw}_3), \nn \\
\partial_z X \sigma_{\theta}(z_0, \zb_0) & \sim & (z-z_0)^{-1+\theta} (z-\zb_0)^{-\theta}, \nn \\
\partial_z \bar{X} \sigma_{\theta}(z_0, \zb_0) & \sim & (z - z_0)^{-\theta} (z - \zb_0)^{-1+\theta},\nn \\
\partial_{\bar{z}} X \sigma_{\theta}(z_0, \zb_0) & \sim & (\bar{z} - z_0)^{-1+\theta} (\bar{z} - \zb_0)^{-\theta},\nn \\
\partial_{\bar{z}} \bar{X} \sigma_{\theta}(z_0, \zb_0) & \sim & (\zb-z_0)^{-\theta} (\zb-\zb_0)^{-1+\theta},
\eea
which follow from the singular behaviour of the bosonic fields on the disk (\ref{OPED} - \ref{OPED3}). We now consider the implications of this.

For the case of a single bulk twist operator we can in fact show that there are no contributions from non-trivial
classical embeddings. On the disk a closed string twist $\sigma_{\theta}(z_1, \zb_1)$  behaves like a twist operator $\sigma_{\theta}$ located at $z_1$ and an anti-twist operator $\sigma_{-\theta}$ located at $z^{\prime}=\bar{z}_1$. The worldsheet field $\partial X(z)$ has a monodromy about $z_1$ and $\bar{z}_1$ given by equations (\ref{yoda2}).
Globally $X(z, \zb)$ takes the form
\be
X(z, \zb) = x_0 + f(z) - f(\bar{z}).
\ee
The monodromy conditions around $z_1$ and holomorphy in $z$
requires $f(z)$ to be such that
\bea
\partial X(z) & = & \alpha(z) (z - z_1)^{-1 + \theta} (z - \bar{z}_1)^{-\theta}, \nn \\
\partial \bar{X} (z) & = & - \alpha^{*}(z) (z - z_1)^{-\theta} (z - \bar{z}_1)^{-1+\theta}.
\eea
The action is
\be
S_{cl} = \frac{1}{4 \pi} \int d^2 z (\partial X_{cl} \bar{\partial} \bar{X}_{cl} + \bar{\partial} X_{cl} \partial \bar{X}_{cl}).
\ee
and we have
\bea
\partial X \bar{\partial} \bar{X} & = & \vert \alpha \vert^2 \vert z - z_1 \vert^{-2 (1-\theta)} \vert z - \bar{z}_1 \vert^{-2 \theta} , \\
\bar{\partial} X \partial \bar{X} & = & \vert \alpha \vert^2 \vert z - z_1 \vert^{-2 \theta} \vert z - \bar{z}_1 \vert^{-2 (1 - \theta)}.
\eea
As $z \to \infty$, both $\partial X \bar{\partial} \bar{X}$ and $\bar{\partial} X \partial \bar{X} \longrightarrow
\vert \alpha \vert^2 \vert z \vert^{-2}$.
If $\alpha(z)$ contains any non-negative powers of $z$ the action is not normalisable as $z \to \infty$.
Likewise, the presence of negative powers of $z$ in $\alpha(z)$ makes the action non-normalisable as $z \to 0$.\footnote{For the case of corrections to gauge kinetic functions, this result is not surprising. Corrections from a non-trivial worldsheet embedding
behave as worldsheet instantons and have contributions to the action that are suppressed as $e^{-2 \pi (t + i b)}$, where $t$ is the volume of a
2-cycle. However for D3/D7 models we know that the appropriate holomorphic K\"ahler moduli are $T_4 + i C_4$ involving 4-cycle volumes complexified
by the RR form. As worldsheet instantons are not holomorphic in these moduli they cannot contribute to the
holomorphic gauge kinetic function.}

This argument does not apply in the case when many twist fields are located at the same orbifold singularity.
If we consider twist fields located at $z_1$, $z_2$, $\ldots$ $z_n$
\be
\partial X(z) \sim (z-z_1)^{-1+\theta} (z-z_2)^{-1+\theta} \ldots (z - z_n)^{-1+\theta} (z - \bar{z}_1)^{-\theta}
(z - \bar{z}_2)^{-\theta} \ldots (z - \bar{z}_n)^{-\theta},
\ee
then we see that the classical action is now normalisable, as $\int \sqrt{g} \partial X \bar{\partial} \bar{X}$ does not diverge
as $z \to \infty$. In this case we expect the presence of non-trivial finite-action classical worldsheet embeddings.

Such a classical solution involves a worldsheet that stretches between the D3-brane stack on the boundary $\hbox{Im}(z) = 0$
and each of the singularities for which a twist field is present, as it must also satisfy $X_{cl}^i(z_1, \zb_1) = x^i_{tw,1}$.
We can use the $SL(2, \mbb{R})$ symmetry of the disk to fix the location of one of the twist fields $(\sigma_{\theta}, \sigma_{-\theta})$
to $(i, -i)$ (this removes any subtleties with twist operators approaching the boundary).
As the worldsheet is constrained to be at $X_i = x_i$ at $\hbox{Im}(z) = 0$ and at $X_i = x^{tw}_i$ at $z=i$, the worldsheet has to stretch.
For generic twist insertions, such a stretched worldsheet has a finite area which scales with the overall radius $R$ of the compact space.
The classical action is simply the area of the worldsheet, leading to a path integral suppression as
$e^{-\lambda \frac{R^2}{2 \pi \alpha'}}$, where $\lambda$ is an $\mc{O}(1)$ number.

There is one important exception to this exponential suppression. Intuitively, the exponential suppression arises because the string is constrained to the orbifold fixed point near the twist insertion. This follows from
the monodromy $X(e^{2 \pi i} z, e^{-2 \pi i} \bar{z}) = e^{2 \pi i \theta / N} X$.
If several twist fields are located on top of each other ($N$ for a $Z_N$ singularity) then the monodromy is eliminated.
In the OPE, bringing several twist fields on top of each other factorises the amplitude into the untwisted sector,
$$
\hbox{lim}_{z_1, z_2 \ldots \to z_n} \sigma_{\theta}(z_1) \sigma_{\theta}(z_2) \ldots \sigma_{\theta}(z_n) = \sum (z - z_n)^{\lambda_i} \mc{O}^{i}_{untw}(z_n).
$$
Untwisted vertex operators do not impose any constraint on the space-time location of the worldsheet and so there is no
exponential suppression in the amplitude.

Given that the boundary condition $X_{cl}(z_n, \zb_n) = x_{tw}(z_n, , \zb_n)$ still holds, it is not at first sight obvious that this eliminates
the exponential suppression, as the worldsheet still has to stretch between the D3 location and the location of the twist fields.
It is easiest to convince oneself that exponential suppression is absent by studying for example
the explicit classical solutions for the four-twist correlation function
found in \cite{Dixon:1986qv} (for example see p45-46 of this paper; the classical action $S_{cl} \sim R^2$ for generic twist locations, but in the
factorisation limit $x \to 1, (\textrm{equivalently } \tau \to 0)$ the classical action vanishes).

\begin{figure}
\begin{center}
\includegraphics[height=5cm]{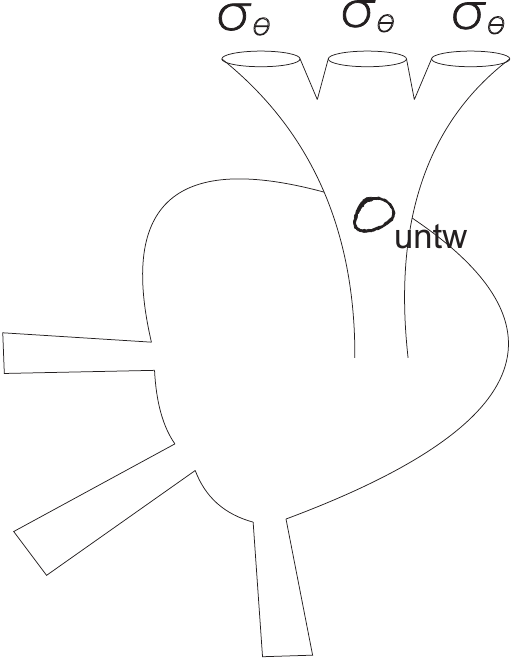}
\caption{The structure of the contributions of distant twist operators to correlation functions. The only non-suppressed contributions
come from regions where the twist operators factorise onto untwisted operators and the amplitude reduces to a correlation function
involving untwisted operators.}
\label{twistfig}\end{center}\end{figure}

In the large radius limit, we can therefore restrict our study of the multi-twist correlation function to the limit where the
twist vertex operators come together and factorise onto states in the untwisted sector, as all other cases are exponentially suppressed
at large radius. This is illustrated in figure \ref{twistfig}.

The correlator (suppressing the antiholomorphic coordinate labels $\zb_i$)
\be
\label{corr}
\langle \sigma_{\theta}(z_1) \sigma_{\theta}(z_2) \ldots \sigma_{\theta}(z_n) C_1 C_2 C_3 \rangle
\ee
gives the S-matrix element for the scattering of $n$ twist fields with the three matter fields $C_1$, $C_2$ and $C_3$.
In the limit $k_i \to 0$, after subtracting reducible lower-point amplitudes this gives terms in the action of the form
\be
\label{dist}
\int \sqrt{-g} \tau_s^n \psi^{SM}_1 \psi^{SM}_2 \phi^{SM}_3,
\ee
which give the dependence of the physical Yukawa couplings on distant blow-up fields. We can use the correlator (\ref{corr})
to deduce the existence or not of a term of the form (\ref{dist}). We do so by subtracting lower point interactions and analysing whether
the residual coupling (\ref{dist}) is present.

We now give a precise argument that terms in the action of the form (\ref{dist}) are absent up to exponentially suppressed
terms of order $e^{-2 \pi R^2}$. The above arguments have already
established that the only case where the classical action for the correlator is not exponentially suppressed is
in the limit where the vertex operators collide,
allowing a factorisation onto the untwisted sector. The factorisation can take two forms: it is either onto massless states in
the untwisted sector (for example the graviton or the dilaton) or it is onto massive states (for example bulk KK modes).

Let us first consider the case of massless modes. In general there is no reason for this not to
lead to a non-zero correlator at zero momentum. For example,
both the dilaton and the graviton have zero-momentum couplings to the Yukawa couplings. The Yukawa couplings depend explicitly on the dilaton (which can be seen by noting that the D3-brane couplings are inherited from N=4 SYM)
$$
\hat{Y}_{\alpha \beta \gamma} = g_s,
$$
while the graviton will couple via the metric interaction
$$
\int \sqrt{g} \psi \psi \phi.
$$
We therefore expect - for example - the worldsheet correlation function $\langle \mc{V}_{\psi} \mc{V}_{\psi} \mc{V}_{\phi} \mc{V}_S \rangle$
to be non-zero at zero momentum.

Factorising onto the dilaton would generate the field theory diagram shown in figure 7. The vertex $\langle \tau_{\theta} \tau_{-\theta} S \rangle$ can only occur at finite momentum (a zero-momentum vertex would correspond to a tree level contribution to the moduli potential, which we know is absent). The diagram then has a positive power of momentum $k_i \cdot k_j$ from the $\tau_{\theta} \tau_{-\theta} S $ vertex and a negative power of momentum $\frac{1}{k_i \cdot k_j}$ from the dilaton propagator, giving overall no powers of momentum.\footnote{We have not analysed whether the $\tau_{\theta}
\tau_{-\theta} S$ vertex actually exists: the point here is to show that such contributions are reducible from the view of the low-energy field theory.}
\begin{figure}
\begin{center}
\includegraphics[height=5cm]{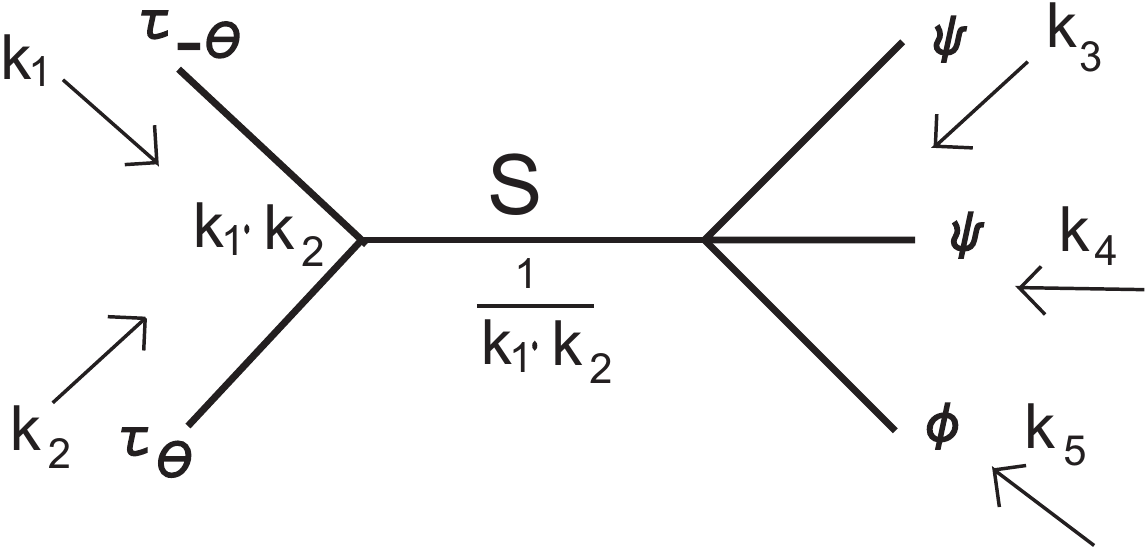}
\caption{The generation of a finite contribution to the correlation function by factoring onto an intermediate massless state such as the dilaton.}
\label{masslessfact}\end{center}\end{figure}

However by construction this interaction is formed simply by gluing together two lower-point interactions that are
already present in the effective field theory. The diagram shown in figure \ref{masslessfact}
can be accounted for by a Lagrangian containing the terms
\be
\mc{L} = S \partial_{\mu} \tau_{\theta} \partial^{\mu} \bar{\tau}_{\theta} + S \psi \psi \phi.
\ee
This in no way implies the existence of an interaction $\tau_{\theta} \tau_{\theta} \psi \psi \phi$ in the low energy effective action.

This argument applies for any factorisation of the twist fields onto massless modes in the untwisted sectors.
In the language of Feynman diagrams this factorisation breaks the original diagrams into two sub-diagrams that are glued together by a propagator of the massless mode.
By construction, the glued diagram involves a lower-point interaction between the twist fields and another massless field.
As the massless fields are necessarily part of the low energy effective field theory, this limit is accounted for by lower-point interactions
and there is no need to include a direct interaction $\tau_{\theta} \tau_{\theta} \ldots \psi \psi \phi$ in the Lagrangian.

The other case is where the factorisation is onto massive modes in the untwisted sector, for example bulk Kaluza-Klein modes.
As the massive states are not in the
low-energy theory, they have to be integrated out and we cannot use the above argument - the Feynman diagram does not
manifestly factorise onto lower-point diagrams in the effective field theory.

In the bosonic string, twist operators have
Yukawa-like interactions with such modes, coming from the OPE
\be
\sigma_{\theta}(z, \zb) \sigma_{-\theta}(w, \wb) \sim (z-w)^h (\bar{z} - \bar{w})^{\bar{h}} e^{i p_L \cdot X_L + i p_R \cdot X_R} , \qquad h, \bar{h} = \half (\frac{m}{R} + \half n R)^2,
\ee
where $p_L, p_R = \frac{m}{R} \pm \frac{nR}{2}$.
These are direct Yukawa interactions with no momentum directly suppressing the 3-point vertex. 

For the superstring the situation is less clear. Direct calculation is beyond the scope of this paper. However 
the twist operators correspond to K\"ahler moduli and so any
Yukawa-like superpotential operator $\tau \tau \Phi_{heavy}$ would violate the perturbative shift symmetry of the K\"ahler moduli (as this 
symmetry originates in the absence of perturbative string couplings to RR modes, it should be unaffected by including KK modes in the action).
This implies that, as with the massless modes, it is not possible to obtain a zero-momentum factorisation onto the heavy KK modes.\footnote{Zero-momentum couplings
of K\"ahler moduli to \emph{pairs} of heavy modes, which are induced from the mass term for the heavy fields through canonical normalisation, 
will give loop diagrams in field theory.} This tells us that the field theory diagram involving factorisation onto a heavy KK modes involves a momentum prefactor coming from the factorisation vertex of twist fields onto KK modes. In principle this factor could be cancelled by an intermediate massless propagator, but as we factorise onto \emph{massive} modes this cannot occur. As a result, the amplitude has an overall momentum prefactor and vanishes at zero momentum, and so cannot give rise to contact terms

We can also give a worldsheet form of this argument,
 through a modified version of an argument in \cite{Dixon:1986qv} (see p60-63), by showing that
this amplitude will necessarily involve a momentum prefactor and thus will vanish at zero momentum. The argument is as follows.
The vertex operators including the twist fields have a canonical $(-1, -1)$ picture. With all fields in the canonical picture,
the correlator we are interested in would be
\be
\langle \sigma^{1}_{(-1,-1)}(w_1) \sigma^{2}_{(-1,-1)}(w_2) \ldots \sigma^{n}_{(-1,-1)}(w_n) \psi_{-1/2}(z_1) \psi_{-1/2}(z_2) \phi_{-1}(z_3) \rangle
\ee
where, for simplicity, we denoted the full twisted vertex operator by $\sigma$.
The disk amplitude must have ghost charge $-2$ and we must picture change appropriately to achieve this.
We picture change the boson $\phi_{-1}(z_3) \to \phi_{0}(z_3)$,
the fermion $\psi_{-1/2}(z_2) \to \psi_{1/2}(z_2)$ and the twist fields $\sigma^{j}_{(-1,-1)}(w_j) \to \sigma^{j}_{(0,0)}(w_j)$ for
$j = 3 \ldots n$.. The first two twist fields are dealt with differently. These are picture changed as
\be
\sigma_{(-1,-1)}(w_1) \to \sigma_{(0,-1)}(w_1), \qquad \sigma_{(-1,-1)}(w_2) \to \sigma_{(0,-1)}(w_2).
\ee
The motivation for this is in the form of the OPE of $e^{\phi} T_F(z) = e^{\phi}( \partial X \cdot \bar{\psi} + \partial \bar{X} \cdot \psi)$
with the twist fields. We have
\bea
\nn \lim_{z \rightarrow w} \quad e^{\phi} \partial \bar{X} \cdot \psi (z) \quad e^{-\phi} e^{-\tilde{\phi}} \sigma_{\theta} s_{\theta} \tilde{s}_{\theta} (w, \wb) & \sim & (z-w), \\
\nn \lim_{z \rightarrow w} \quad e^{\phi} \partial X \cdot \bar{\psi} (z) \quad e^{-\phi} e^{-\tilde{\phi}} \sigma_{\theta} s_{\theta} \tilde{s}_{\theta} (w, \wb) & \sim & 1, \\
\nn \lim_{\zb \rightarrow \wb} \quad e^{\tilde{\phi}} \bar{\partial} \bar{X} \cdot \tilde{\psi} (\zb) \quad e^{-\phi} e^{-\tilde{\phi}} \sigma_{\theta} s_{\theta} \tilde{s}_{\theta} (w, \wb) & \sim & 1, \\
\lim_{\zb \rightarrow \wb} \quad e^{\tilde{\phi}} \bar{\partial} X \cdot \bar{\tilde{\psi}} (\zb) \quad e^{-\phi} e^{-\tilde{\phi}} \sigma_{\theta} s_{\theta} \tilde{s}_{\theta} (w, \wb) & \sim & (\zb-\wb).
\label{xylop}
\eea
The significance of the OPEs is that internal picture changing of the operator $\sigma_{\theta}$ can only be done via the element
$\partial X \cdot \bar{\psi}$ of $T_F(z)$ (and in particular cannot be done with $\bar{\partial} X \cdot \psi$). This introduces 2 units of
positive H charge. By picture changing both $\sigma_{-1,-1}(w_1)$ and $\sigma_{(-1,-1)}(w_2)$ in this way, 4 units of internal positive H-charge
are introduced (if we instead picture
changed $\sigma_{(-1,-1)}(w_1) \to \sigma_{(-1,0)}(w_1), \sigma_{(-1,-1)}(w_2) \to \sigma_{(-1,0)}(w_2)$ four units
of negative H-charge would be introduced). This H-charge cannot be cancelled through purely internal picture changing: the twists $\sigma^3 $ to $\sigma^n$ involve both $\partial X \cdot \bar{\psi}$ and $\bar{\partial} \bar{X} \cdot \tilde{\psi}$, and so give no net H-charge.\footnote{One could attempt to cancel the H-charge against the picture changing of the open string fields. However one can check that this cannot work, as the open string
boson can only contribute to one sign of the H-charge. By an appropriate choice of picture changing either to $(0,-1)$ or $(-1,0)$, we can ensure that it is not possible to obtain compensating H-charge from the open string fields.} The result is that the contribution from purely internal picture changing necessarily vanishes.

The significance of this is that picture changing of the twist operators $\sigma_{-1,-1}(w_1)$ and $\sigma_{-1,-1}(w_2)$ must involve the
external directions. This involves two contractions of $\partial X$ with a momentum operator $e^{ik \cdot X}$, giving a
kinematic invariant $k_i \cdot k_j$ as a prefactor. This factor can only be cancelled by a propagator for an intermediate massless field,
 but since we have already been able to restrict to factorising onto massive fields this cannot occur.
As a result the overall amplitude has a kinematic prefactor $k_i \cdot k_j$, and so the amplitude vanishes at zero momentum.

The conclusion of this argument it that at string tree level there is no term perturbative in $\alpha'$ in the
(canonically normalised) effective action of the form
\be
\label{lab}
\int \sqrt{g} \tau_s \tau_s \ldots \tau_s \psi \psi \phi,
\ee
where $\tau_s$ corresponds to a blow-up mode located at a different singularity than that of the D3-brane.
Although the argument has been developed in terms of twist fields located at a single singularity, it extends
without modification to the case of many twist fields located at multiple singularities. In this case we equip the twist fields with a label indicating the orbifold singularity. The results of this section can then be directly applied to all twist fields carrying the same singularity label.

The argument had three steps:
\begin{enumerate}
\item
For generic locations of twist operators, the classical worldsheet has to stretch between them.
The classical action $e^{-S_{cl}}$ is then exponentially suppressed unless the
twist operators coincide in a factorisation limit.
In this limit we can use the OPE to factor the twist operators onto either massless or massive states in the untwisted sector.
\item
If the untwisted states are massless, we can decompose the diagram into sub-diagrams that come from lower point interactions in the
effective field theory. These lower point interactions are sufficient for this limit and there is no need for a term of
the form (\ref{lab}).
\item
If the states are massive, the picture changing can be chosen to show that the amplitude has a kinematic prefactor that vanishes at zero momentum, excluding a term of the form (\ref{lab}).
\end{enumerate}
The only possible loophole in the worldsheet argument would seem to be if there was a reason why the picture changing described below
eq. (\ref{xylop}) should be forbidden.

While these calculations have been carried out in the orbifold limit, the importance of the multi-twist results
lie in the fact they allow us to extend the results beyond the orbifold limit. If the blow-up modes $\tau_s$ acquire vevs, then the
orbifold is resolved to become a smooth Calabi-Yau space. The absence of the Lagrangian terms (\ref{lab}) shows that even when
we resolve onto the smooth space, there is still no dependence of the physical Yukawa couplings on the blow-up modes (up to terms
non-perturbative in $\alpha'$).\footnote{Strictly this results holds only up the radius of convergence of the expansion in $\tau_s$.}
This applies for the resolution of either one or many of the singularities that are present.

The significance of this is that it shows that, at leading order in $g_s$ and to all perturbative orders in $\alpha'$, the physical Yukawa couplings
\be
\hat{Y}_{\alpha \beta \gamma} = e^{K/2} \frac{Y_{\alpha \beta \gamma}}{\sqrt{Z_{\alpha} Z_{\beta} Z_{\gamma}}},
\ee
do \emph{not} depend on the blow-up moduli $\tau_s$.
This is in addition to the independence of the physical Yukawa couplings from the bulk K\"ahler moduli, which follows
automatically from the orbifold model (we can quantise the string and compute the $\langle \psi \psi \phi \rangle$ correlator
irrespective of the size of the bulk tori). This ties subleading $\alpha'$ corrections to $Z$ to the subleading $\alpha'$ corrections
to $K$ in such a way that the overall physical Yukawa couplings remain independent of the K\"ahler moduli.

For IIB flux compactifications there is a subleading correction to the K\"ahler potential coming from $\alpha'^3$ corrections, \cite{bbhl}
\be
K = - 2 \ln \mc{V} \to K + \delta K = - 2 \ln \left( \mc{V} + \frac{\xi}{(S + \bar{S})^{3/2}} \right).
\ee
Despite the appearance of the dilaton, this correction is tree level in $g_s$: the presence of the dilaton comes from the
definition of the chiral supergravity variables. The above result then confirms a conjecture of \cite{09063297} that the correction
to the matter metrics would be such as to cancel the correction to the moduli K\"ahler potential when considering the physical Yukawas.

We note that this is modulo the assumption that the equivalence of $\sqrt{Z_{\alpha} Z_{\beta} Z_{\gamma}}$ and $e^{K/2}$
also leads to the equivalence of $Z_{\alpha}$ and $e^{K/3}$. If this is not the case then would need to examine individual formulae
carefully to see which cancellations hold and which do not.

The significance of the structure $Z \sim e^{K/3}$ in relation to the K\"ahler moduli is that this is precisely the condition necessary for the
cancellation of many contributions to soft terms that are present for models of supersymmetry breaking based on no-scale supersymmetry breaking,
including the models of \cite{0502058, 0505076}. This condition was also called 'sort-of sequestering' in \cite{10121858}. For the LARGE volume scenario this leads to a cancellation in soft terms at $\mc{O}(m_{3/2})$
and $\alpha_i \mc{O}(m_{3/2})/4 \pi$, as well as the cancellation of certain contributions at $\mc{O}(m_{3/2}^{3/2} M_P^{-1/2})$.

\section{Conclusions}

In this paper we have studied the sequestering of local D-brane constructions from distant moduli fields.
We have done so by computing the dependence of physical Yukawa couplings on the blow-up modes
that resolve singularities. In the LARGE volume models these blow-up modes correspond to the
small cycles that are necessary for moduli stabilisation and supersymmetry breaking, and have non-zero
F-terms.

The heart of the paper is the computation of correlators between matter fields and twist moduli. This involves an
extension of the formalism of \cite{Dixon:1986qv} for twist vertex operators to the disk. We have checked this formalism through
computations involving a twist field and two gauge bosons or two matter fields, reproducing the expected results.
The main calculation has then been that of three open string vertex operators corresponding to the $\psi \psi \phi$ Yukawa coupling
together with closed string twist fields $\sigma_{\theta}$. This calculation gives the dependence of physical Yukawa couplings on the vevs
of twist moduli. This is important as, modulo a mild assumption, this is equivalent to `sort-of sequestering' of twist moduli and a
resulting suppression in the scale of soft terms.

For the case of a single twist field we have determined the full quantum and classical correlator. In this case the classical correlator is
easy, being non-normalisable if the twist field is at a different singularity to the matter sector and trivial if the twist field is at the same
singularity. The quantum correlator gives the full behaviour of the 4-point function for arbitrary momenta of the external fields. By analysing
the structure at zero momentum, we determine the presence of a contact term $\int {\textrm{d}}^4 x \ \tau_s \psi \psi \phi$ in the effective action. This shows explicitly that, as expected, twist fields at the same singularity are not sequestered.

For the case of multiple twist fields the full quantum correlator is beyond the scope of this paper. However by a judicious choice of picture
changing we were able to establish the absence of contact terms $\int {\textrm{d}}^4 x \ \tau_s \tau_s ... \tau_s \psi \psi \phi$ in the effective action when
$\tau_s$ corresponds to a singularity distant from the matter fields. This was done by showing that the multi-twist correlator was exponentially
suppressed except in a factorisation limit, and any terms present from the OPE in the factorisation limit were either reducible to lower point amplitudes or came with prefactors of momentum and so vanished at zero momentum. Although the calculations are performed in the orbifold
limit, by giving vevs to the twist fields $\tau_s$ we can extend these results to smooth Calabi-Yaus. As the resolution of toroidal orbifold
singularities leads to a space resembling the Swiss-cheese structure required for a LARGE volume scenario, this provides
a method to obtain CFT results for the geometries appropriate to the LARGE volume scenario.

Finally, the results obtained in this paper come from a disk calculation in CFT. This gives results that are valid at all orders in $\alpha'$ but only at leading order in $g_s$. Various subtle effects can occur at 1-loop level \cite{10030388, 10110999, 11085740} and one very interesting extension of this work would be to repeat the same calculation on the annulus, to see whether the results found here persist at the loop level.

\subsection*{Acknowledgments}
We thank Marcus Berg, Mark Goodsell, David Marsh, Liam McAllister and Eran Palti for inspirational discussions. JC is supported by a Royal Society University Research Fellowship and Balliol College. LW is supported by the Science and Technology Facilities Council (STFC).

\appendix

\section{Integrals}
\label{theusefulintegral}
An integral that plays an important role here is
\be
\label{GarousiMyersInt}
I(a, b, c, d, e, f) = (2i)^f \int_{-\infty}^{\infty} dx_1 \int_{x_1}^{\infty} dx_2
(x_1 - i)^a (x_1 + i)^b (x_2 - i)^c (x_2 + i)^d (x_2 - x_1)^e.
\ee
It evaluates to
\bea
\label{GManalytic}
I(a,b,c,d,e,f) & = & -(2i)^{3+a+b+c+d+e+f} \Gamma(-2-a-b-c-d-e) \ti \nn \\
& & \Big[ (-i)^{2(a+c)} \frac{\sin \pi(b+d+e) \Gamma(1+e) \Gamma(2+b+d+e) \Gamma(-1-d-e)}{\Gamma(-d)\Gamma(-a-c)} \ti \nn \\
& & \phantom{.}_3 F_2(-c,1+e,2+b+d+e;2+d+e,-a-c;1) \nn \\
& & + (-i)^{2(a+c+d+e)} \frac{\sin (\pi b) \Gamma(-1-c-d-e) \Gamma(1+b) \Gamma(1+d+e)}{\Gamma(-c) \Gamma(-1-a-c-d-e)} \ti \nn \\
& & \phantom{.}_3 F_2(-d,-1-c-d-e,1+b;-d-e,-1-a-c-d-e;1) \Big]
\eea
The analytic evaluation can be found in \cite{Garousi:1996ad} and for completeness we review it here. The correctness of this expression can be checked
by numerically evaluating the integral (\ref{GarousiMyersInt}) and comparing to the analytic expression (\ref{GManalytic}).

The first part is to perform the integration over $x_1$. This can be carried out using
equation (3.197)(1) of \cite{Gradshteyn},
\be
\int_0^{\infty} x^{\nu-1} (x+\beta)^{-\mu} (x+\gamma)^{-\rho} = \beta^{-\mu} \gamma^{\nu - \rho} B(\nu, \mu - \nu + \rho)
\phantom{.}_2 F_1 \Big(\mu, \nu;\mu + \rho; 1- \frac{\gamma}{\rho} \Big).
\ee
This leaves
$$
(2i)^f \int_{-\infty}^{+\infty} dx_2 \, (x_2 - i)^{a+c} (x_2 + i)^{b+d+e+1} B(e+1, -c-e-1-d) \phantom{.}_2
F_1 \left(-c,e+1;-c-d;1-\frac{x_2 + i}{x_2 -i}\right).
$$
We next use relation (9.131)(2) from \cite{Gradshteyn},
\bea
\phantom{.}_2 F_1(\alpha, \beta, \gamma; 1- z) & = & \frac{\Gamma(\gamma)\Gamma(\gamma - \alpha - \beta)}{\Gamma(\gamma - \alpha)
\Gamma(\gamma - \beta)} \phantom{.}_2 F_1(\alpha, \beta; \alpha + \beta - \gamma + 1; z)  \\
& & + z^{\gamma - \alpha - \beta} \frac{\Gamma(\gamma) \Gamma (\alpha + \beta - \gamma)}{\Gamma(\alpha) \Gamma(\beta)}
\phantom{.}_2 F_1(\gamma - \alpha, \gamma - \beta; \gamma - \alpha - \beta + 1; z). \nn
\eea
This gives
\bea
(2i)^f \int_{-\infty}^{+\infty} dx_2 \, (x_2 - i)^{a+c} (x_2 + i)^{b+d+e+1} \frac{\Gamma(e+1)\Gamma(-d-e-1)}{\Gamma(-d)}
\phantom{.}_2 F_1(-c, e+1 ; d+e+2; x_2 - i)  & & \nn \\
+ (x_2 - i)^{a+c+d+e+1} (x_2 + i)^b \frac{\Gamma(-c-d-e-1)\Gamma(d+e+1)}{\Gamma(-c)}
\phantom{.}_2 F_1(-d,-c-d-e-1; -d-e; \frac{x_2 +i}{x_2 - i}). & & \nn
\eea
Next we use the explicit expansion of the hypergeometric function
\be
\phantom{.}_2 F_1(\alpha_1, \alpha_2; \beta; z) = \frac{\Gamma(\beta)}{\Gamma(\alpha_1) \Gamma(\alpha_2)}
\sum_{n=0}^{\infty} \frac{\Gamma(\alpha_1 + n) \Gamma(\alpha_2 + n)}{\Gamma(\beta + n)} \frac{z^n}{n!}.
\ee
This gives
\bea
& & (2i)^f \frac{\Gamma(1+d+e)\Gamma(-d-e)}{\Gamma(-c) \Gamma(-d)} \ti  \\
& & \Bigg( \int_{-\infty}^{+\infty} dx_2 \, -(x_2 - i)^{a+c-n} (x_2 + i)^{b+d+e+1+n} \sum_{n=0}^{\infty} \frac{\Gamma(-c+n)\Gamma(e+1+n)}
{\Gamma(d+e+2+n) n!} \nn \\
& & + \int _{-\infty}^{+\infty} dx_2 \, (x_2 - i)^{a+c+d+e+1-n} (x_2 + i)^{b+n} \sum_{n=0}^{\infty} \frac{\Gamma(-d+n)\Gamma(-c-d-e-1+n)}{
\Gamma(-d-e+n)n!} \Bigg) \nn
\eea
We now use the identities
\bea
\int_{-\infty}^{+\infty} dx \, (x-i)^A (x+i)^B & = & \frac{-\pi (-i)^{2A} (2i)^{2+A+B} \Gamma(-1-A-B)}{\Gamma(-A)\Gamma(-B)}, \\
\Gamma(-A)\Gamma(1+A) & = & \frac{- \pi}{\sin(\pi A)}, \\
\frac{\Gamma(1+a+c-n)}{\Gamma(-1-b-d-e-n)} & = & \frac{-\sin \pi(b+d+e) \Gamma(2+b+d+e+n)}{\sin \pi(a+c) \Gamma(-a-c-n)}, \\
\frac{\Gamma(2+a+c+d+e-n)}{\Gamma(-b-n)} & = & \frac{-\sin (\pi b) \Gamma(1+b+n)}{\sin \pi(a+c+d+e) \Gamma(-1-a-c-d-e+n)}.
\eea
These enable us to write $I(a,b,c,d,e,f)$ as
\bea
& & (2i)^{3+a+b+c+d+e+f} \frac{\pi \Gamma(-2-a-b-c-d-)}{\sin \pi(d+e) \Gamma(-c) \Gamma(-d)} \ti \\
& & \Bigg( \sum_{n=0}^{\infty} -(i)^{2(a+c)} \sin \pi(b+d+e) \frac{\Gamma(-c+n)\Gamma(1+e+n)\Gamma(2+b+d+e+n)}{n! \Gamma(2+d+e+n) \Gamma(-a-c+n)} \nn \\
& & + (-i)^{2(a+c+d+e)} \sin \pi b \frac{\Gamma(-d+n) \Gamma(-1-c-d-e+n) \Gamma(1+b+n)}{n! \Gamma(-d-e+n) \Gamma(-1-a-c-d-e+n)} \Bigg)
\eea
Using the definition of $\phantom{.}_3 F_2$,
\be
\phantom{.}_3 F_2 (\alpha_1, \alpha_2, \alpha_3; \beta_1, \beta_2; 1) =
\frac{\Gamma(\beta_1)\Gamma(\beta_2)}{\Gamma(\alpha_1)\Gamma(\alpha_2)\Gamma(\alpha_3)}
\sum_{n=0}^{\infty} \frac{\Gamma(\alpha_1 + n)\Gamma(\alpha_2 + n) \Gamma(\alpha_3 + n)}{\Gamma(\beta_1 + n)\Gamma(\beta_2 + n) n!},
\ee
we now obtain the result (\ref{GManalytic}).

\section{Result for other cyclic ordering}
In the main text we only considered orderings of the open strings on the boundary that are equivalent to $z_1 z_2 z_3$ under cyclic permutations. For a given Yukawa coupling we only need to calculate the amplitudes for one cyclic ordering as the Chan-Paton factors vanish for the other cyclic ordering. Nevertheless, for completeness we also present the amplitudes for orderings of the open strings on the boundary that are equivalent under cyclic permutations to $z_2 z_1 z_3$:
\bea
\mc{A}(z_2z_1z_3)_{z_3 \to \infty} &: & (u+t) \ti I \Big(\theta_2 + t/2, 1 - \theta_2 + t/2, -1+ \theta_1 + u/2, - \theta_1 + u/2, -1 + s, -1 \Big) \nn \\
& & +\frac{t}{2} \ti I \Big(-1+ \theta_2 + t/2, 1 - \theta_2 + t/2, -1+ \theta_1 + u/2, - \theta_1 + u/2, s, -1 \Big) \nn \\
& & +\frac{t}{2} \ti I \Big(\theta_2 + t/2, - \theta_2 + t/2, -1+ \theta_1 + u/2, - \theta_1 + u/2, s, -1 \Big) \\
& & -I \Big(\theta_2 + t/2, 1 - \theta_2 + t/2, -1+ \theta_1 + u/2, - \theta_1 + u/2, -1 + s, -1 \Big) \nn \\
& & + 2i \theta_3 \ti I \Big(\theta_2 + t/2, - \theta_2 + t/2, -1+ \theta_1 + u/2, - \theta_1 + u/2, -1 + s, -1 \Big). \nn \\
& & \nn \\
\mc{A}(z_1z_3z_2)_{z_2 \to \infty} & : & \frac{t}{2} \ti I \Big(\theta_1 + u/2, - \theta_1 + u/2, \theta_3 +s/2, -\theta_3+s/2, -1+t, -1 \Big) \nn \\
& & +\frac{t}{2} \ti I\Big(-1 + \theta_1 + u/2, 1 - \theta_1 + u/2, \theta_3 +s/2, -\theta_3+s/2, -1+t, -1 \Big) \nn \\
& & - u \ti I \Big(-1 + \theta_1 + u/2, - \theta_1 + u/2, \theta_3 +s/2, -\theta_3+s/2, t, -1 \Big) \\
& & I \Big(-1 + \theta_1 + u/2, - \theta_1 + u/2, \theta_3 +s/2, -\theta_3+s/2, t, -1 \Big) \nn \\
& & + 2 i \theta_3 I\Big(-1 + \theta_1 + u/2, - \theta_1 + u/2, -1+ \theta_3 +s/2, -\theta_3+s/2, t, -1 \Big). \nn \\
& & \nn \\
\mc{A}(z_3z_2z_1)_{z_1 \to \infty} & : & \frac{t}{2} \ti I \Big(\theta_3 + s/2,1 - \theta_3 + s/2, -1+\theta_2 +t/2, 1-\theta_2+t/2, -1+u, -1 \Big) \nn \\
& & +\frac{t}{2} \ti I\Big(\theta_3 + s/2,1 - \theta_3 + s/2, \theta_2 +t/2, -\theta_2+t/2, -1+u, -1 \Big) \nn \\
& & +u \ti I \Big(\theta_3 + s/2,1 - \theta_3 + s/2, \theta_2 +t/2, 1-\theta_2+t/2, -2+u, -1 \Big) \\
& & - I \Big(\theta_3 + s/2,1 - \theta_3 + s/2, \theta_2 +t/2, 1-\theta_2+t/2, -2+u, -1 \Big) \nn \\
& & -2i\theta_3 I\Big(-1+\theta_3 + s/2,1 - \theta_3 + s/2, \theta_2 +t/2, -\theta_2+t/2, -1+u, -1 \Big). \nn
\eea

\section{Pole structure of amplitudes}
\label{poles}
Here we collect the results for the poles in the partial amplitudes arising in the calculation of a Yukawa coupling with a twist insertion. The results were derived by expanding the full result for the amplitudes for small momenta. In some cases we were able to expand the generalized hypergeometric function $_3F_2$ appearing in the expression for the amplitude analytically; when this was not possible, the following results were obtained numerically and can be checked to hold for arbitrary angles $\theta_i$ for $\theta_1+\theta_2+\theta_3=1$.
\begin{align}
&\mc{A}_{z_3 \to \infty} =  \\
& \frac{i e^{\pi i \theta_3} (-1+4\theta_3) \sin(\pi \theta_3)}{s}  + \frac{i e^{-\pi i \theta_1+\pi i \theta_3} \hphantom{-}(1-2\theta_2) \sin(\pi \theta_2)}{t} + \frac{i e^{-\pi i \theta_1} (-1+2\theta_1-2\theta_3) \sin(\pi \theta_1)}{u} \nn \\
& \vphantom{blank} \nn \\
& e^{-2 \pi i (\theta_1+\theta_2)} \mc{A}_{z_2 \to \infty} = \\
&  \frac{i e^{\pi i \theta_3} (-1+2\theta_3) \sin(\pi \theta_3)}{s} \hphantom{+ \ } \hphantom{\frac{i e^{-\pi i \theta_1+\pi i \theta_3} (-1-2\theta_2) \sin(\pi \theta_2)}{t}} + \frac{i e^{-\pi i \theta_1} (-2\theta_3) \sin(\pi \theta_1)}{u} \nn \\
& \vphantom{blank} \nn \\
& -e^{-2 \pi i \theta_1}\mc{A}_{z_1 \to \infty} = \\
&  \frac{i e^{\pi i \theta_3} \hphantom{-}(1-2\theta_3) \sin(\pi \theta_3)}{s} + \frac{i e^{-\pi i \theta_1+\pi i \theta_3} (-1+2\theta_2) \sin(\pi \theta_2)}{t} + \frac{i e^{-\pi i \theta_1} (1-2\theta_1) \sin(\pi \theta_1)}{u} \ . \nn
\end{align}
We need to sum these partial results to arrive at the full amplitude. The expressions with poles in $t$ cancel when added. Once we modify the expression $e^{-2 \pi i (\theta_1+\theta_2)} \mc{A}_{z_2 \to \infty} \rightarrow - e^{-2 \pi i (\theta_1+\theta_2)} \mc{A}_{z_2 \to \infty}$ we see that the poles in $u$ also sum to zero. The remaining pole in $s$ is the result given in \eqref{Afullpole}.

\bibliographystyle{JHEP}

\bibliography{TwistJCLWv2}

\end{document}